\documentclass[12pt,preprint]{aastex}

\usepackage{graphicx}
\usepackage{amssymb}
\usepackage{epstopdf}
\usepackage{ulem}

\usepackage{natbib}
\newcommand{\chandra}{{\it CHANDRA}}
\newcommand{\rxte}{{\it RXTE}}

\newcommand{\xmm}{{\it XMM}}

\newcommand{\ec}{$\eta$~Carinae}

\slugcomment{Accepted by the Astrophysical Journal}

\shorttitle{X-ray Flaring from $\eta$ Car}
\shortauthors{Moffat \& Corcoran}

\begin{document}


\title{Understanding the X-ray Flaring from \ec}


\author{A. F. J. Moffat}
\affil{D\'epartement de physique, Universit\'e de Montr\'eal, Succursale Centre-Ville, Montr\'eal, QC, H3C 3J7, and Centre de recherche en astrophysique du Qu\'ebec, Canada}
\email{moffat@astro.umontreal.ca}

\author{M. F. Corcoran\altaffilmark{1}}
\affil{CRESST and X-ray Astrophysics Laboratory NASA/GSFC, Greenbelt, MD 20771, USA} 
\email{michael.f.corcoran@nasa.gov}




\altaffiltext{1}{Universities Space Research Association, 10211 Wincopin Circle, Suite 500 Columbia, MD 21044, USA.} 

\begin{abstract}
We 
quantify the rapid variations in X-ray brightness (``flares'') from the extremely massive colliding wind binary \ec\ 
seen during the past three orbital cycles by \rxte.  
The observed flares tend to be shorter in duration and more frequent as periastron is approached, although the largest ones tend to be roughly constant in strength at all phases.  Plausible scenarios include (1) the largest of multi-scale stochastic wind clumps from the LBV component entering  
and compressing the hard X-ray emitting wind-wind collision (WWC) zone, (2) large-scale corotating interacting regions in the LBV wind sweeping across the WWC zone, or (3) instabilities intrinsic to the WWC zone. The first one appears to be most consistent with the observations, requiring homologously expanding clumps as they propagate outward in the LBV wind and a turbulence-like power-law distribution of clumps, decreasing in number towards larger sizes, as seen in Wolf-Rayet winds. 
\end{abstract}




\section{Introduction}

The extremely massive binary star \ec\ consists of a Luminous Blue Variable (LBV) primary (star A) coupled with a hot, fast-wind secondary (star B) which is probably an evolved O star \citep{2005ApJ...624..973V} or a Wolf-Rayet star \citep{2002AA...383..636P}. Orbital and stellar properties gleaned from the literature are given in Table \ref{tab:params}. These values are considered to be the best current values, although overall relatively uncertain. Star A has gone through at least one giant LBV eruption, which occurred in the 1840s \citep{1997ARA&A..35....1D}.

\begin{deluxetable}{lcl}
\tablecaption{Adopted Parameters for the \ec\ Binary System}
\tablehead{
\colhead{Parameter} & \colhead{Value} & \colhead{Reference}
}
\startdata
$P$(orbit) & $2024.0 \pm 2$ d = $5.541 \pm 0.006$ yr & \cite{2005AJ....129.2018C}\\
T$_\circ$\tablenotemark{a} & HJD $2450800\pm3$ d (=1997.9618 UT) & \cite{2005AJ....129.2018C}\\
$i$ & $45^{\circ}$ &  \cite{2008MNRAS.388L..39O}\\
$e$ & 0.9 &  \cite{2008MNRAS.388L..39O}\\
$\omega$ & 243$^{\circ}$  &  \cite{2008MNRAS.388L..39O}\\
$a$ & 15.4 AU & \cite{2001ApJ...547.1034C}\\
$D$(periastron) & 1.54 AU & \\
$D$(apastron) & 29.3 AU & \\
$M_{A}$ & 90  M$_{\odot}$ & \cite{2001ApJ...553..837H}\\
$M_{B}$ & 30 M$_{\odot}$ & \cite{2005ApJ...624..973V}\\
${\dot M}_{A}$ & 10$^{-3}$ M$_{\odot}$ yr$^{-1}$ & \cite{2001ApJ...553..837H}\\
${\dot M}_{B}$ & 10$^{-5}$ M$_{\odot}$ yr$^{-1}$ & \cite{2002AA...383..636P}\\
$V_{\infty,A}$ & 500 km/s & \cite{2001ApJ...553..837H}\\
$V_{\infty,B}$& 3000 km/s &  \cite{2002AA...383..636P}\\
$R_{A}$ & 0.28 AU & \cite{2001ApJ...553..837H}\\
\enddata
\tablenotetext{a}{T$_{\circ}$ is the Heliocentric Julian Day Number at the start of the X-ray minimum in 1997-1998, which is likely close to the time of periastron passage.}
\label{tab:params}
\end{deluxetable}%

\ec\ generates X-ray emission in the $2-10$ keV range due to the collision of star A's dense, slow wind with the thinner, faster wind of star B \citep{1999ApJ...524..983I}.  On timescales of years, the X-ray emission varies in a manner similar to that expected from a long-period, highly eccentric, colliding-wind binary, with $L_X \propto 1/D$ (where $D$ = orbital separation), for adiabatic conditions as is most likely the case in such a long-period system \citep{usov92}, though near periastron passage radiative effects \citep{2009MNRAS.tmp..279P} may become important. This trend is broken by a broad atmospheric eclipse in the X-ray light-curve, which 
begins near periastron passage when the dense inner wind of Star A starts to block out most of the X-ray emission arising in the bow-shock head region of the wind-wind collision (WWC) zone. Along with this long-term variation, short term variations in \ec's X-ray brightness, or ``flares'' \citep{1997Natur.390..587C,1998NewA....3..241D,1999ApJ...524..983I}, have been observed by \rxte\ for more than two full orbital cycles from 1996.2 up to at least 2009.0, near the date of this writing.  These flares are not expected in standard colliding wind models, and their origin is currently unknown.  We examine here the characteristics of these flares and discuss possible physical mechanisms for their production.  

\section{The \rxte\ Observations}

The Rossi X-ray Timing Explorer \citep[\rxte,][]{1993A&AS...97..355B} is a satellite X-ray observatory designed to provide high time-resolution, low spectral resolution observations of bright X-ray sources.  For our purposes we restrict ourselves to discussion of data from \rxte's Proportional Counter Array (PCA), which consists of five collimated Proportional Counter Units (PCUs 0--4) sensitive to emission in the $2-60$ keV band with a field of view of  $\sim1^{\circ}$ FWHM. \rxte\ was launched in December 1995 and began observations of \ec\ in February 1996. 

The \rxte\ observations in the first two cycles of \ec\ were described in \citet{2005AJ....129.2018C}, with similar observing parameters for the most recent data.  Each PCA observation 
lasts typically 1--2 ksec and is usually obtained at a cadence of a few times per month. For certain key times in the cycle (for example, just before the 2003 X-ray minimum and during the decline to minimum in 2008-2009)  daily X-ray observations were obtained.  Prior to the 1997-1998 minimum, observations were done weekly, with daily observations only after the beginning of the X-ray minimum.  Obtaining frequent observations just prior to X-ray minima is important to temporally resolve the flaring behavior of the source, especially as the X-ray maximum is approached when the flare durations tend to be much shorter. 
%
%
The data were extracted from PCU2 since it is the only Proportional Counter Unit which has continuously observed \ec\ from the start of the campaign and thus the PCU2 data provide the most uniform set of observations available. We reduced the data as described by \citet{2005AJ....129.2018C} and references therein. In addition to those steps, we also reprocessed all the data since 1996 to correct for newly-discovered problems in the SAA (South Atlantic Anomaly) history file that result in errors in background rate estimation, along with a bug in the \rxte\ background estimator (PCABACKEST) for faint models in Epoch 5c (corresponding to data obtained from 01 Jan 2004 to the present). 

\section{The Residual Flare-Lightcurve}

Characterizing the flares is not easy, especially for most of the cycle 
away from the X-ray minimum when the flares are broad and weak.  Flaring occurs on top of a slowly-varying phase-dependent baseline brightness-level produced by the shocked gas in the WWC, which (as noted above) mostly varies as $1/D$, where $D$ is the instantaneous separation between the centers of the two stars. In addition there can be small-scale variations in the slow baseline from one observation to the next because of uncertainties in the correction for instrumental background or, possibly, intrinsic variations in the source (or variability from another source in the PCA field of view).  While Poisson-based errors are typically $\sim$0.1 ct/s, background uncertainties can reach as high as $\sim$1 ct/s rms.

\begin{figure}[htbp] 
   \centering
   \includegraphics[width=6in]{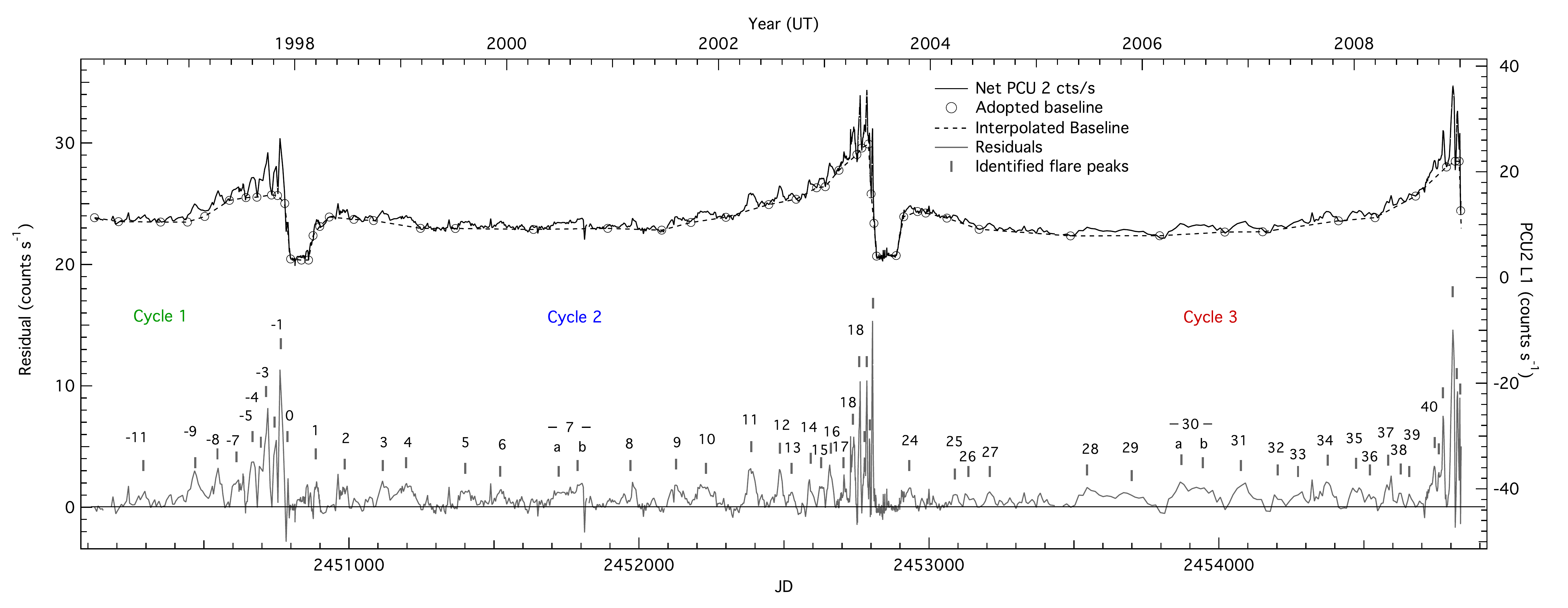} 
   \includegraphics[width=2in]{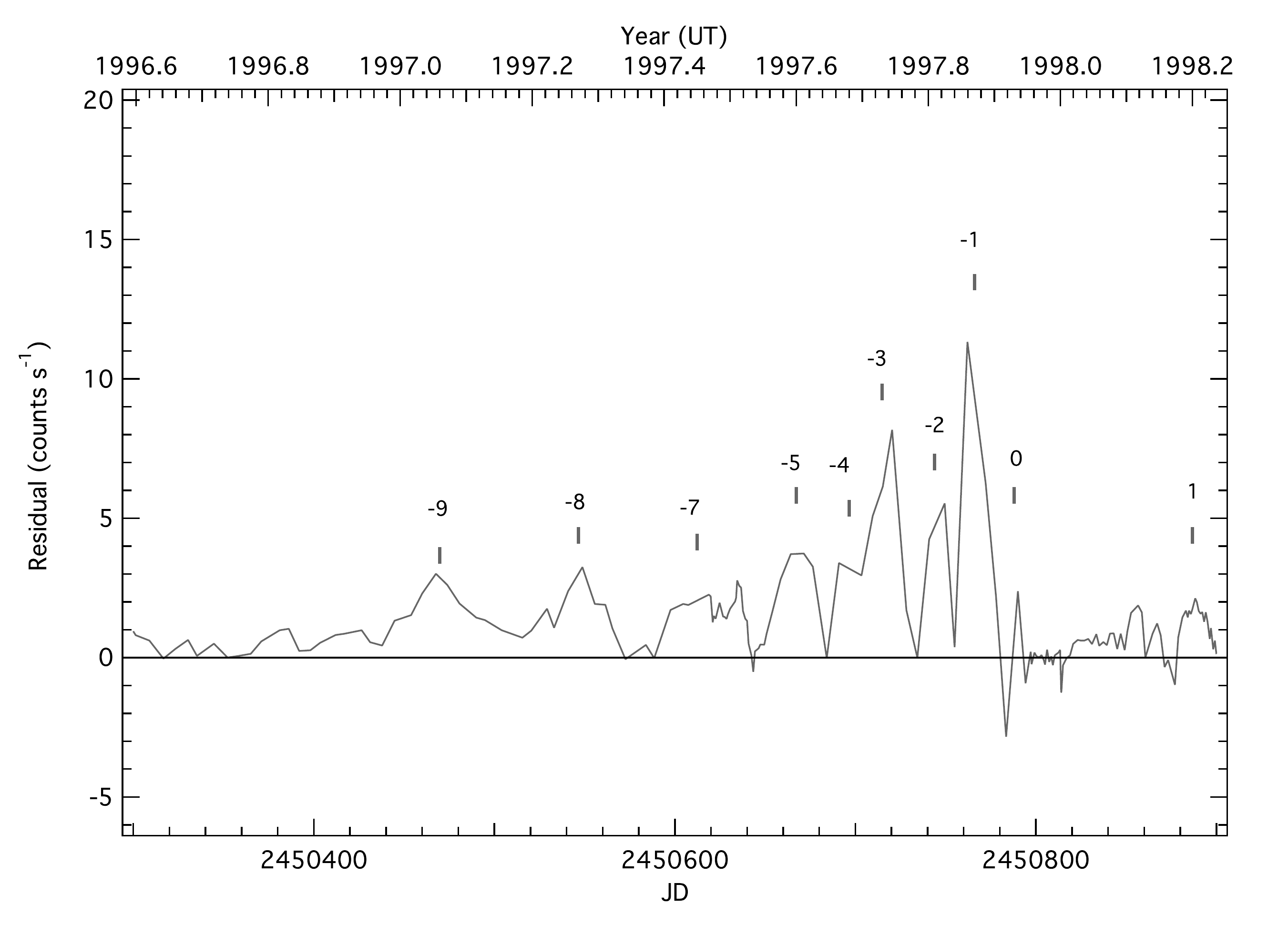} 
   \includegraphics[width=2in]{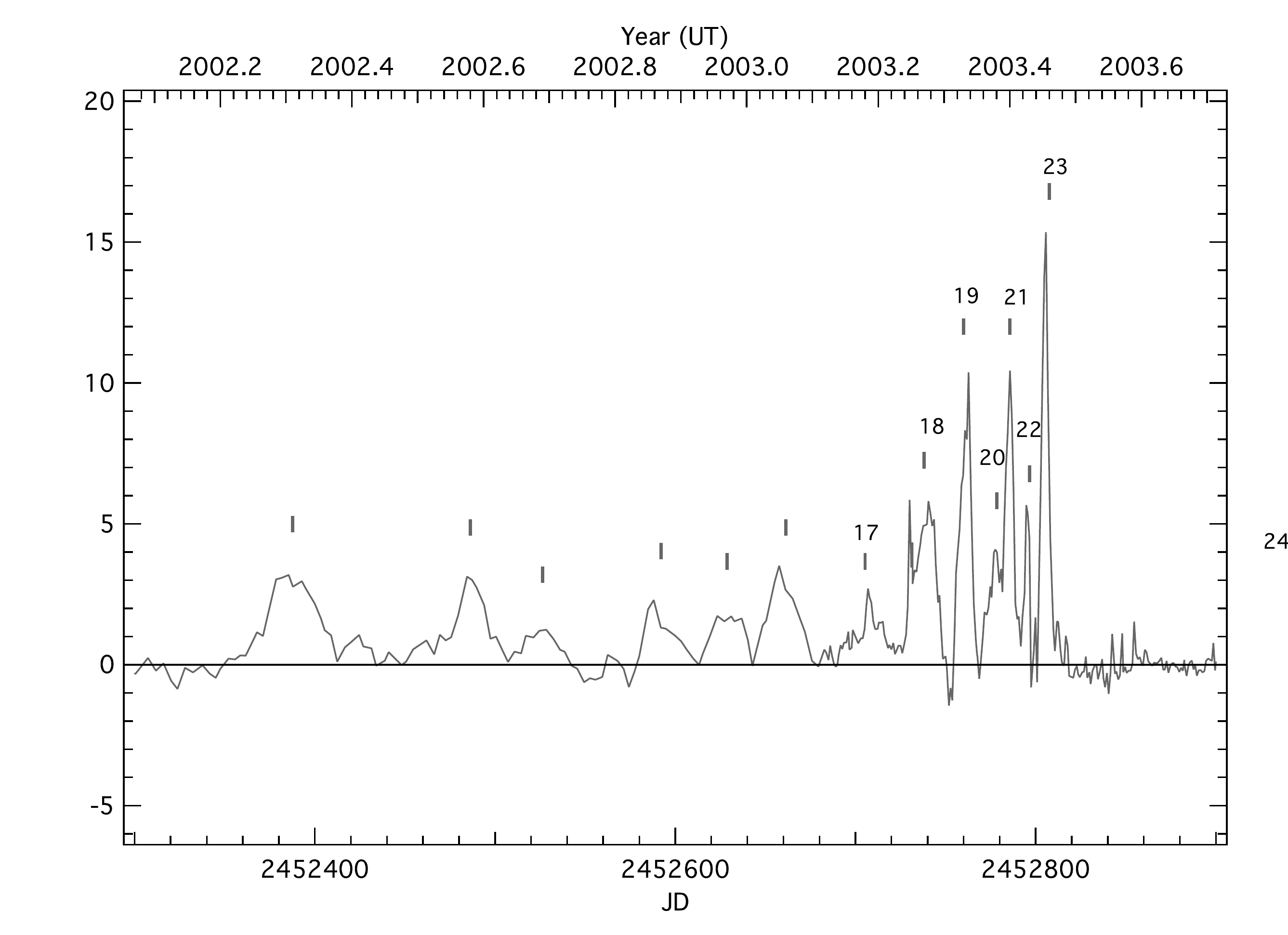} 
   \includegraphics[width=2in]{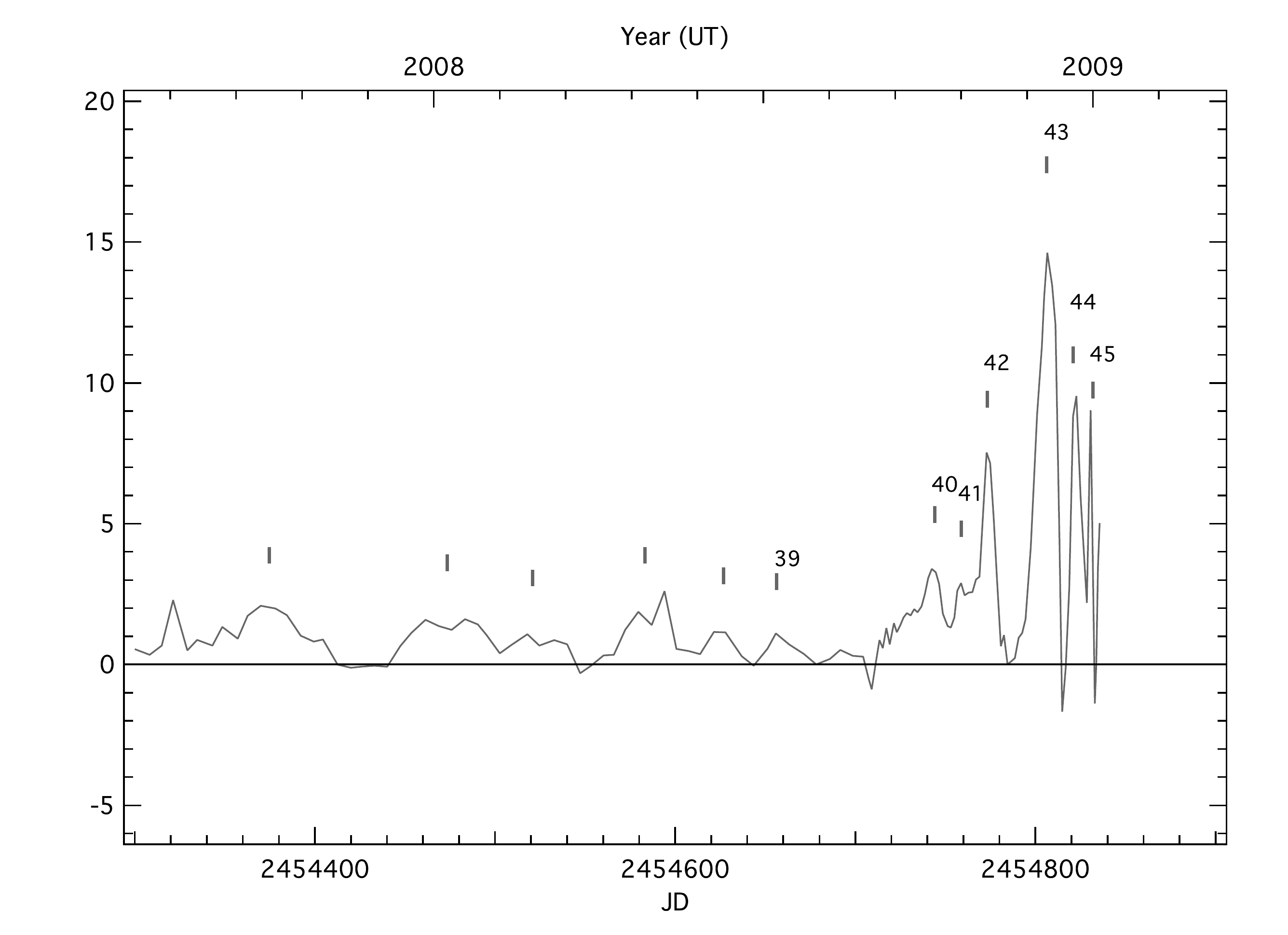} 
   \caption{Top: \rxte\ X-ray lightcurve (solid black line), adopted brightness level for underlying (baseline) X-ray variations (dashed line), and residuals (solid gray line).  The filled circles show the points adopted to represent the underlying brightness due to the colliding-wind emission. Tick marks indicate the positions of the peaks of the identified flares, along with the flare identifications from Table \ref{tab:flareparams} (except just prior to minima when the flare spacing is too small). Bottom:  Zoomed plots of the strong flare intervals around the 3 X-ray minima; tick marks show flare peaks along with the flare ID numbers from Table \ref{tab:flareparams}.}
   \label{fig:noflareamp}
\end{figure}

\begin{figure}[htbp]
\begin{center}
\includegraphics[width=6in]{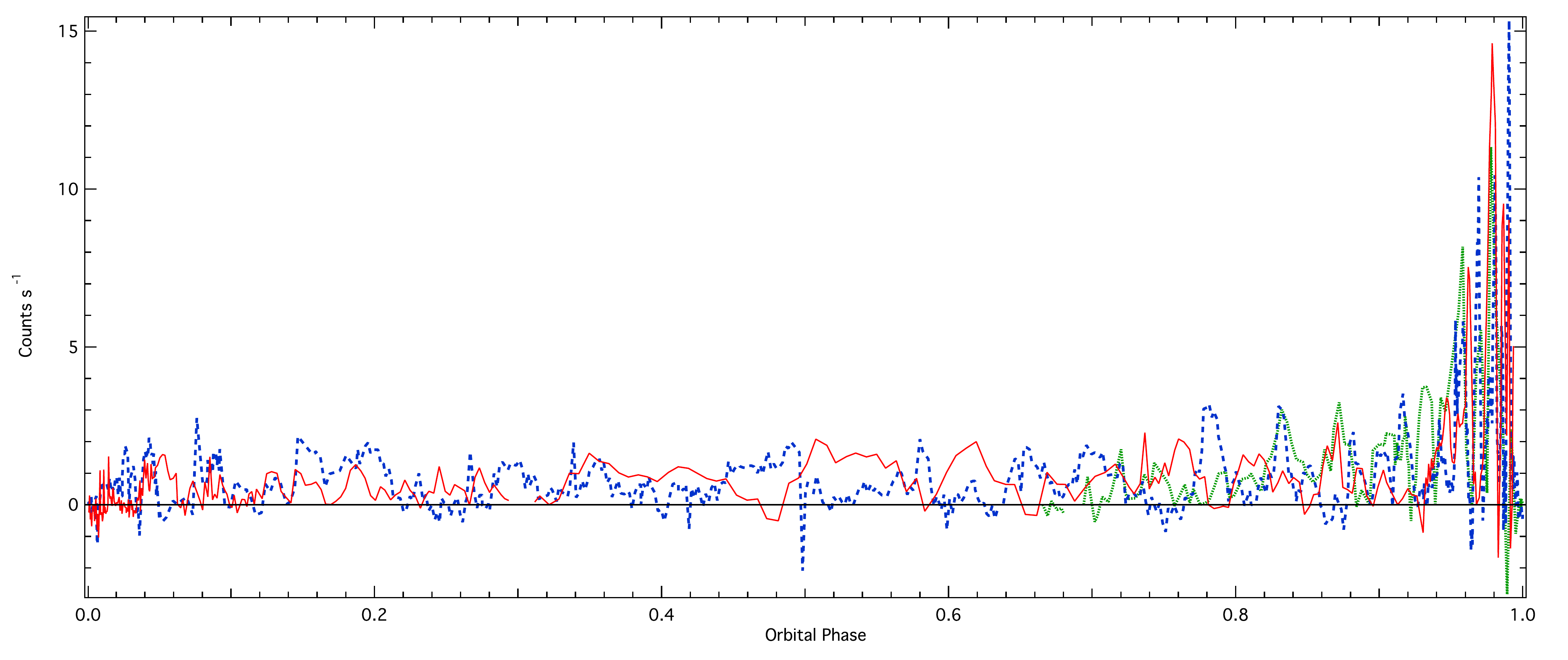}
\includegraphics[width=6in]{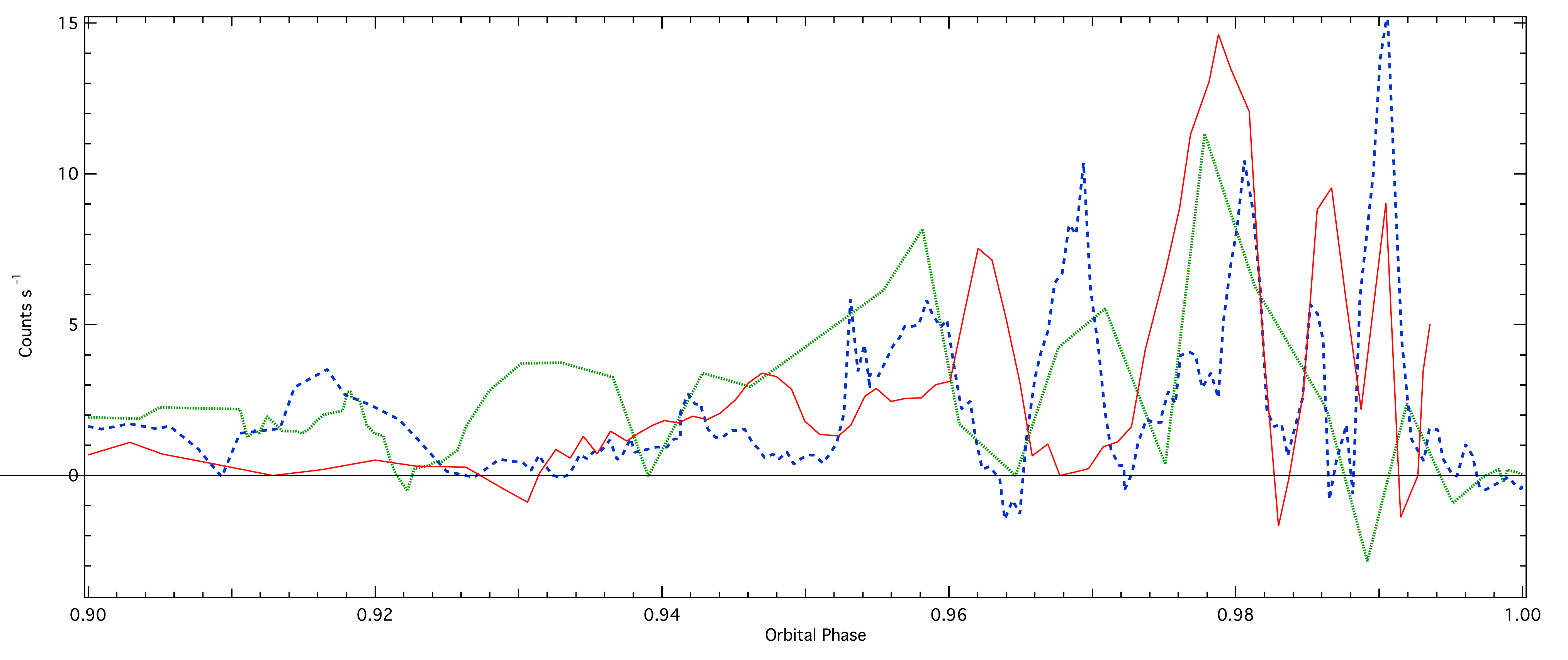}
\caption{Top: Comparison of residual net rates (after subtracting the adopted ``baseline'') for the three cycles seen by \rxte. Green is cycle 1, blue cycle 2, and red cycle 3. Bottom: Close-up of phase interval $0.9<\phi<1.0$.}
\label{fig:residuals}
\end{center}
\end{figure}

We define a ``residual flare-lightcurve'' as the height 
of the observed PCU2 net count rate above the variable-baseline colliding-wind emission in the following way.  We first selected by eye a set of 55 data points to represent the apparent non-flare baseline brightness level due to the underlying colliding-wind emission.  Using these data points, we then created a piecewise continuous function to represent the underlying (non-flare) brightness of the source by linearly interpolating the discrete data points to the times of the full \rxte\ lightcurve.  We then subtracted this piecewise continuous curve from the background-corrected \rxte\ lightcurve to generate the ``residual flare-lightcurve''.  Figure \ref{fig:noflareamp} shows the discrete data points we selected along with the piecewise continuous brightness variation, and the residual ``flare lightcurve'' after subtracting the interpolated baseline.
We divide the entire residual ``flare lightcurve'' into 3 cycles by phase.  Cycle 1 refers to phases $-0.35 < \phi < 0.00$, cycle 2 refers to $0.00 < \phi < 1.00$, and cycle 3 refers to $1.00<\phi<2.00$, where phases $\phi$ are calculated as 
\begin{equation}
\phi=(t - T_{o})/P,
\label{eq:ephem}
\end{equation}
where $t$ is the Heliocentric Julian day number of the start of the observation and $T_{o}$ and $P$ are given in Table 1. Figure \ref{fig:residuals} compares the flares in a phase plot (see below for discussion).

\begin{figure}[htbp]
\begin{center}
\includegraphics[width=6in]{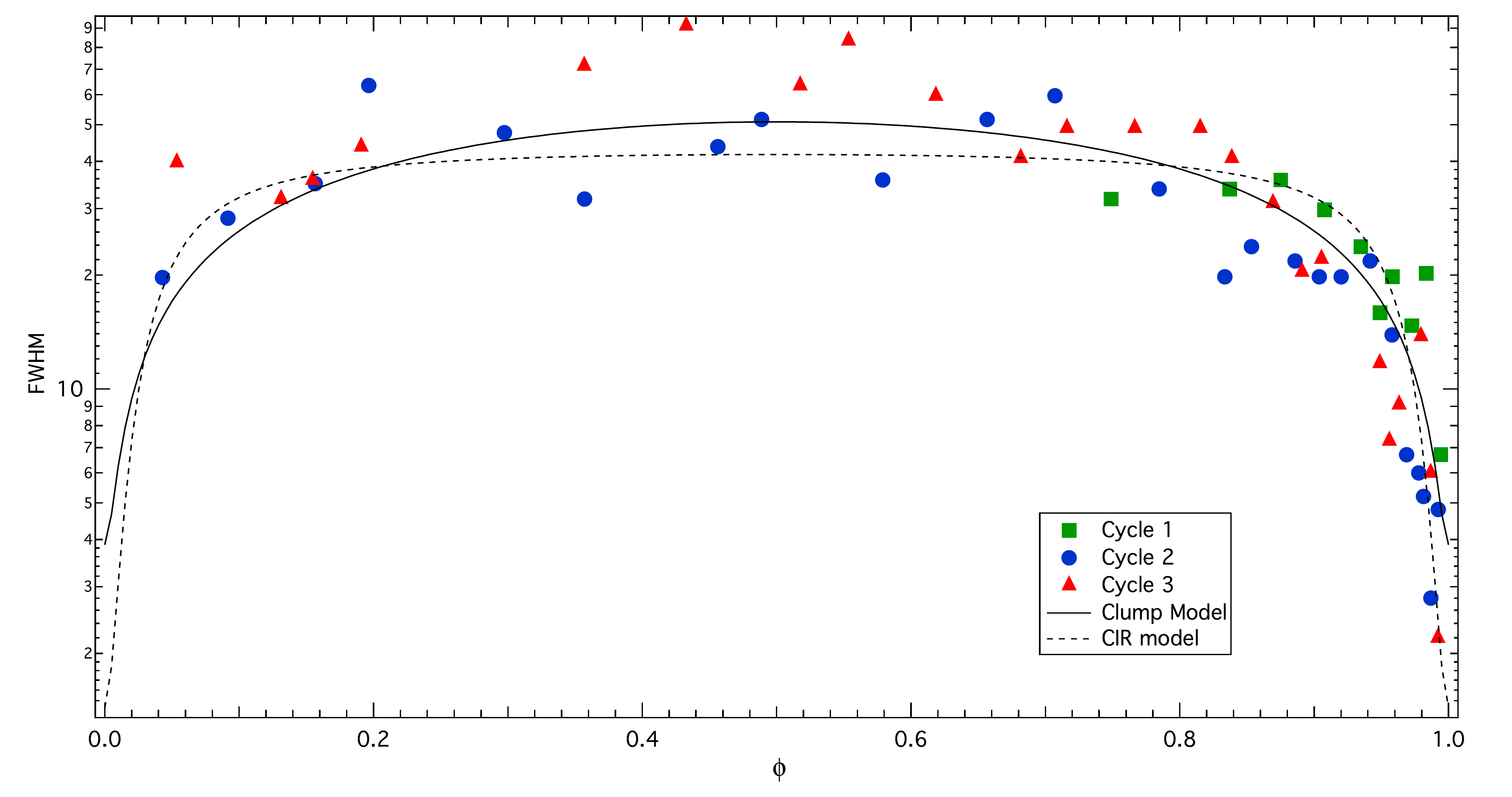}
\caption{Full width half maximum (in days) for the identified flares listed in Table \ref{tab:flareparams} vs. orbital phase. Green symbols are from cycle 1, blue symbols cycle 2, and red symbols cycle 3. The smooth curves are the best-fit models described in the text: clump model (see eq. \ref{eq:clumpfwhm}), long-dashed line; CIR model, short dashed line.}
\label{fig:fwhm}
\end{center}
\end{figure}
\begin{figure}[htbp]
\begin{center}
\includegraphics[width=6in]{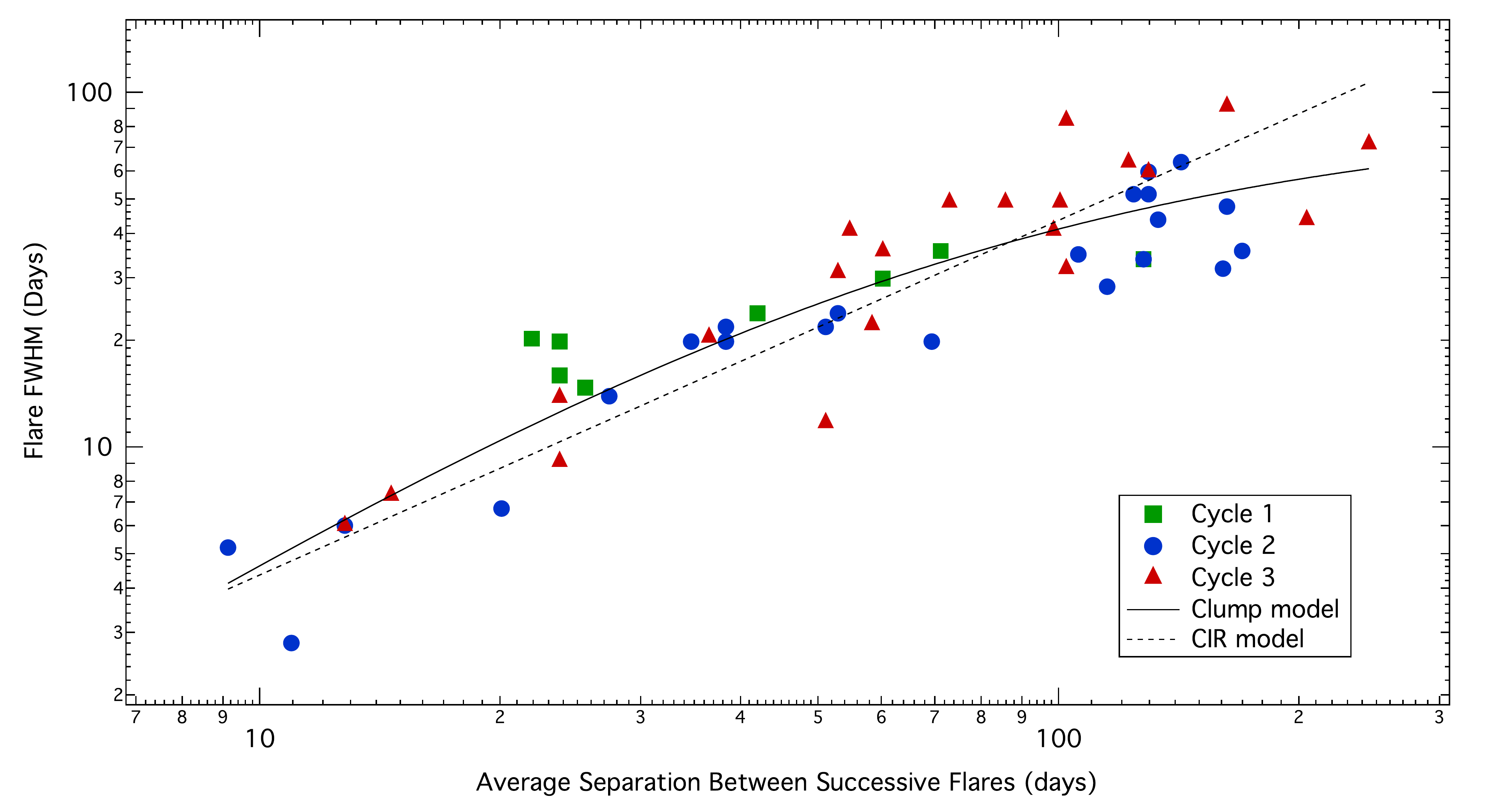}
\caption{Flare FWHM vs. average separation between successive flares. Green symbols are from cycle 1, blue symbols cycle 2, and red symbols cycle 3, along with the best clump-model fit and the best fit CIR model.}  
\label{fig:tm_flare_fwhm_ave_sep}
\end{center}
\end{figure}
\begin{figure}[htbp]
\begin{center}
\includegraphics[width=6in]{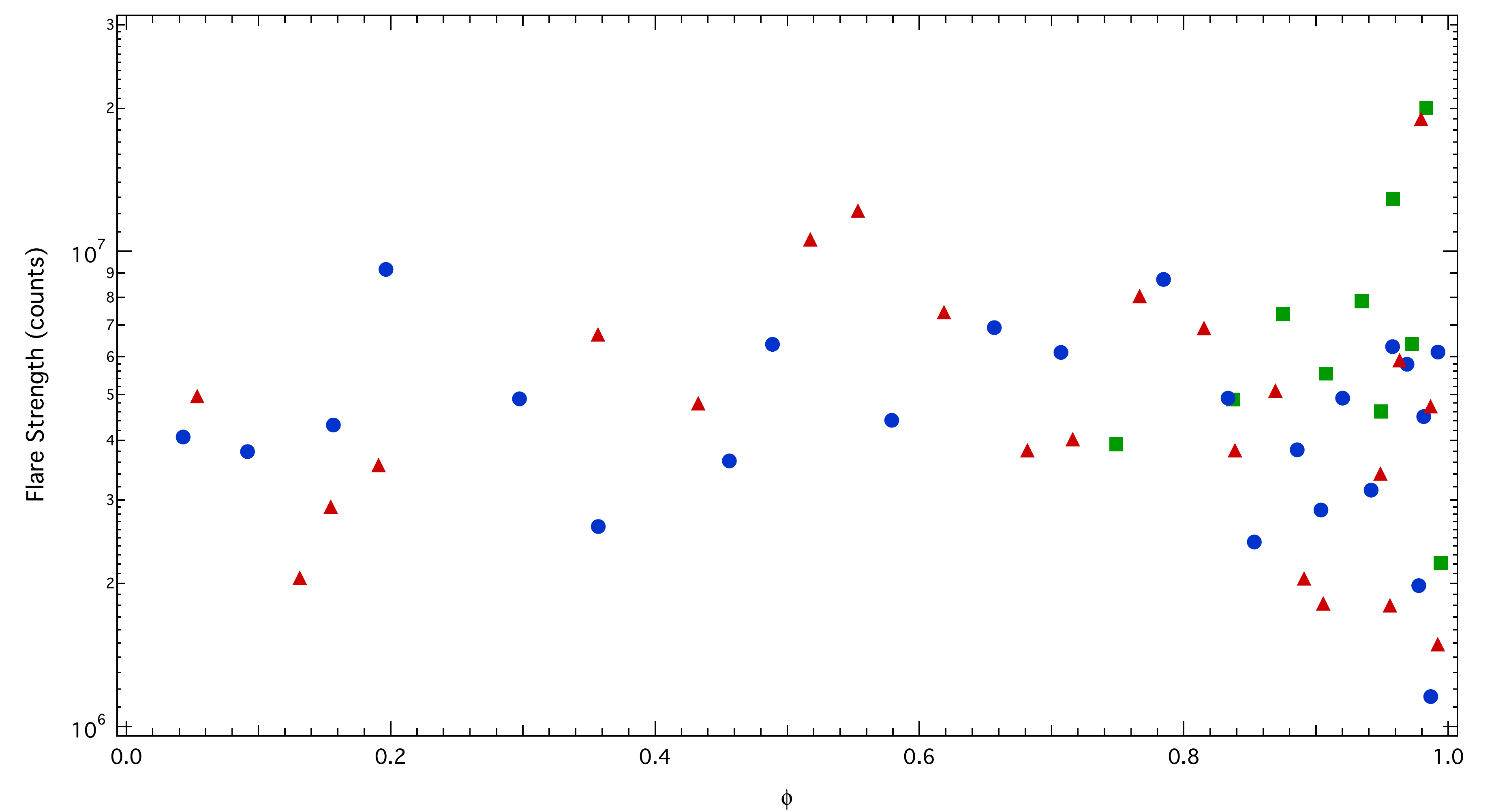}
\caption{Plot of flare strength (peak $\times$ FWHM) vs. phase.}
\label{fig:strength}
\end{center}
\end{figure}

Flares which occur at higher X-ray baseline flux just before plummeting to X-ray minimum are relatively narrow and intense and thus easy to identify.  At lower baseline X-ray fluxes, flares are weak and broad, and determining flare parameters for them is more subjective.   
After selection by eye of all the most obvious flares whose peak value lies at least $\sim$1 ct/s above the zero level of the residual flare-lightcurve, we then measured the flare properties: time corresponding to the flare centroid, flare temporal full width 
at half maximum (FWHM), and peak flare height above the baseline (H).  Table \ref{tab:flareparams} (appendix) lists the measured quantities for each identified flare.
Figure \ref{fig:fwhm} 
shows FWHM as a function of phase for each of the three cycles.  Time separation for a given flare (also given in Table \ref{tab:flareparams}) was obtained by averaging the time interval from the prior flare to the current flare and the time interval from the current flare to the next flare. 
Figure \ref{fig:tm_flare_fwhm_ave_sep} shows FWHM versus time separation. 
Both quantities are clearly correlated, although not exactly linearly.  Figure \ref{fig:strength} shows a phase plot of the total strength $S$ of the flares, estimated as $S = \ $FWHM $\times H$. 
We see that, within the scatter, the maximum  total strength is $\sim$independent of phase, whereas one sees a larger number of weaker flares closer to periastron.  Clearly, all three cycles show the same trends with phase in all these quantities (FWHM, separation and strength).

%
%

\subsection{Comparison of Identified Flares}
\label{sec:flarecmp}

\begin{figure}[htbp]
\begin{center}
\includegraphics[width=6in]{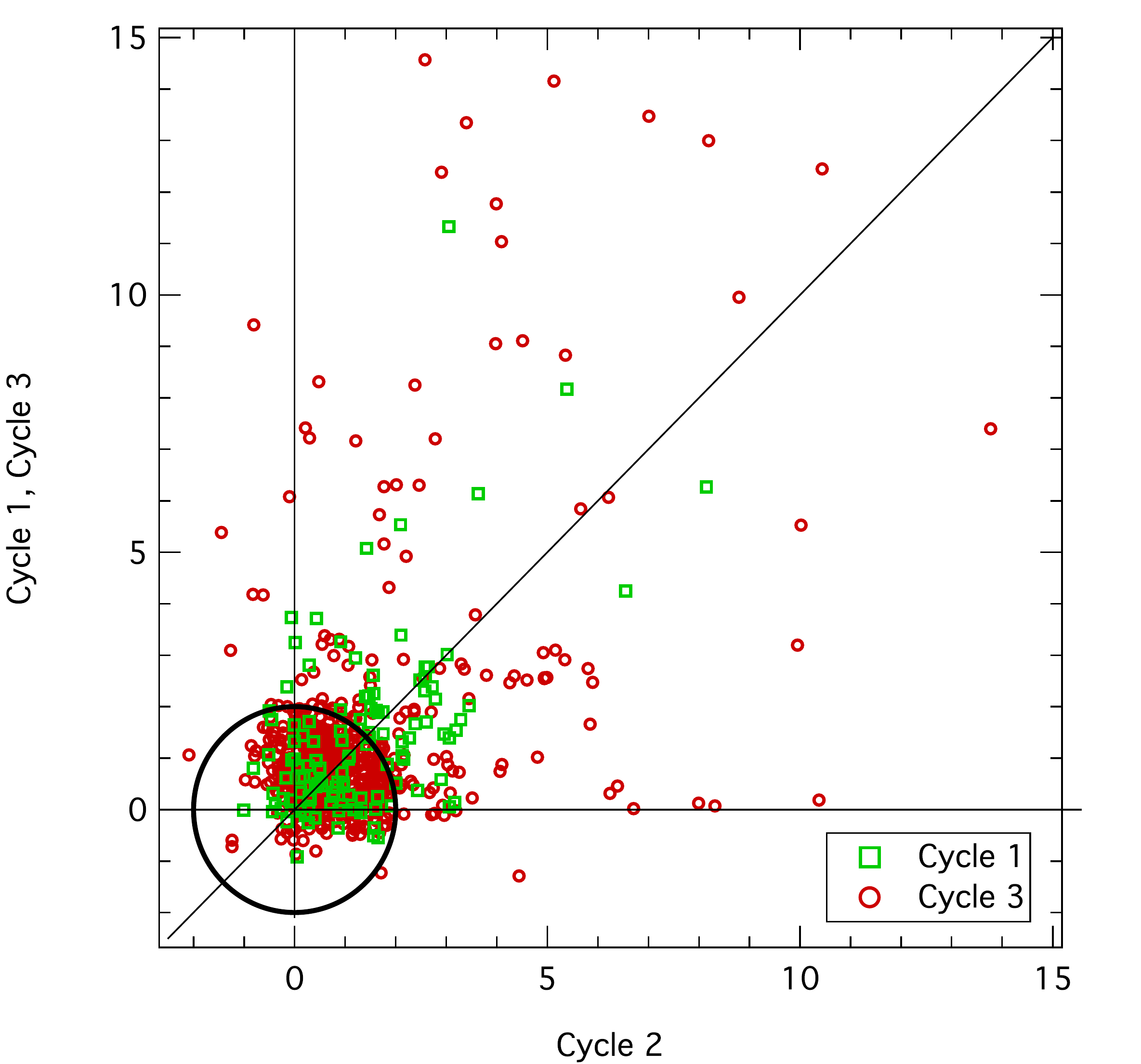}
\caption{Plot in residual intensity units (counts/s) of cycle 1 (green squares) and cycle 3 (red circles) 
vs. cycle 2.  We interpolated the phasing of the cycle 3 curve to that of cycle 2 before plotting.  In order to avoid spurious artifacts, we interpolated the cycle 2 data to the (poorer-sampled) cycle 1 phasing before plotting. The solid circle is centered on (0,0) with a radius of 2 counts/s, below which random errors dominate.}
\label{fig:cycle-comparisons}
\end{center}
\end{figure}

Figure  \ref{fig:residuals} compares the whole residual flare lightcurves in the three different cycles, including a close-up just before the X-ray minimum. 
There is no apparent detailed repetition of the flares from one cycle to the next, although a few of the flare phasings seem to repeat in two (or in rare cases, all three) cycles. Based on a limited sample of the Cycle 1 data,  \cite{1997Natur.390..587C} and \cite{1999ApJ...524..983I} suggested a periodicity of $\approx 85$ days for the X-ray flares which they suggested might arise in stellar pulsations of \ec~A. 
We here attempt 
a more objective test for repeatability of the flares from one cycle to the next now using data from all 3 cycles observed by \rxte.  For this purpose, Figure \ref{fig:cycle-comparisons} first compares the complete residual lightcurves of cycles 1 and 3 to that of cycle 2.  Because the sampling rates are different, for this comparison we interpolated the  cycle 3 residuals to the phasing of the cycle 2 data; because cycle 1 is somewhat undersampled compared to cycle 2 (and cycle 3), we interpolated the phasing of the cycle 2 data to the phasing of the cycle 1 data, then overplotted the cycle 1 residuals vs. the interpolated cycle 2 residuals.  In Figure \ref{fig:cycle-comparisons}, both axes are biased to positive values (making a simple correlation analysis inappropriate), except for data points close to the origin, which are dominated by noise (and systematic uncertainty in our determination of the baseline level).  

\begin{figure}[htbp]
\begin{center}
\includegraphics[width=6in]{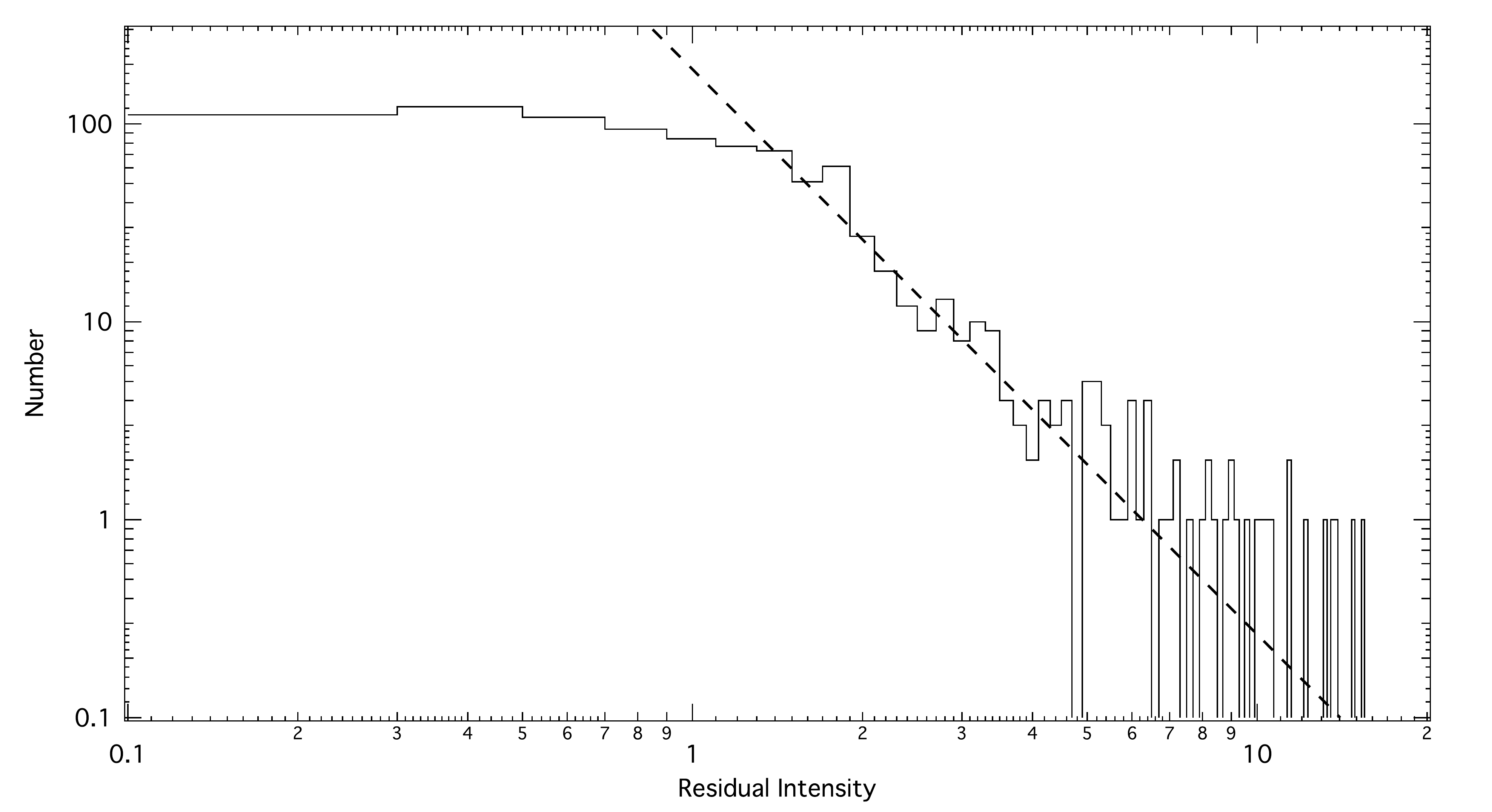}
\caption{Histogram of residual flare intensity ( i.e. all intensity points after subtracting baseline).  Turnover occurs at an intensity of about 1.3 cts s$^{-1}$.  Above this value the distribution is characterized by a power-law of slope $-2.86\pm0.31$ as shown by the dashed line.} 
\label{fig:residuals-histogram}
\end{center}
\end{figure}

Figure \ref{fig:residuals-histogram} shows a histogram of the residual intensities for all 3 cycles.  Here, we see a trend of constant slope only above a critical value $I_c \sim$ 1.3 cts/s, to which we fit a straight line out to $I = $ 10 cts/s (above which the bins become too sparsely populated).  Below $I_c$ the values flatten out;  we take this as the residual count-rate limit below which the data are noise dominated. We note that the slope of the straight-line fit in Figure \ref{fig:residuals-histogram} is $b=-2.9\pm0.3$.  This slope is somewhat steeper than that found for the variable components of optical emission lines in WR stars: $b\approx-2.0\pm0.2$ on average 
\cite{1994Ap&SS.221.....M}, where we may be seeing the power-law spectrum due to full-scale, supersonic compressible turbulence in their winds.

\begin{figure}[htbp]
\begin{center}
\includegraphics[width=6in]{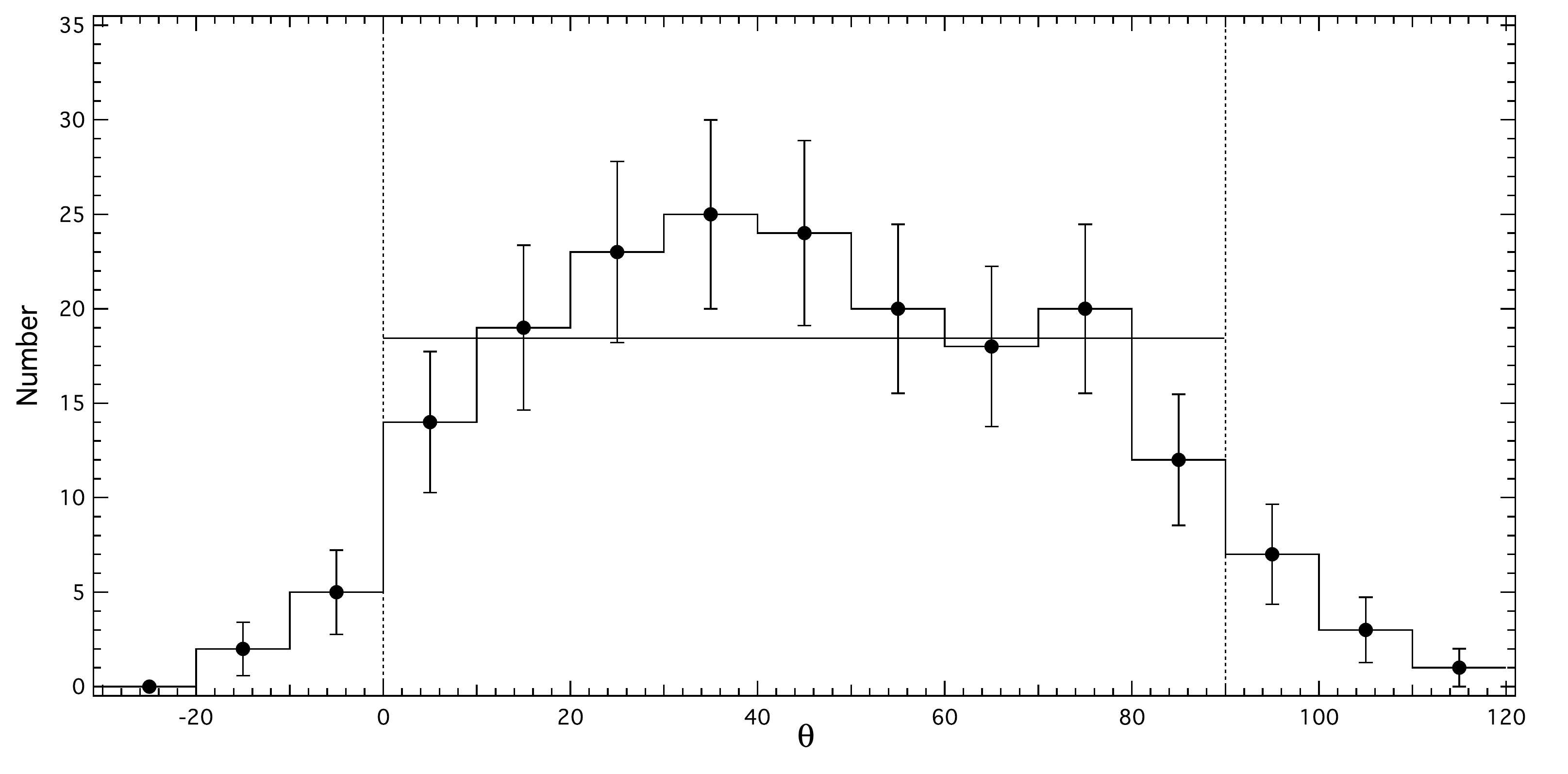}
\caption{Histogram of the distribution in polar angle of the residual intensity (see 
figure \ref{fig:cycle-comparisons}). The horizontal line is a constant mean value between $0^{\circ}$ and $90^{\circ}$, which is an adequate description of the distribution in this angular range ($\chi^{2}_{\nu}= 1.01$).}
\label{fig:residuals-histogram-theta}
\end{center}
\end{figure}

Figure \ref{fig:residuals-histogram-theta} shows the variation with polar angle $\theta = \arctan(x/y)$ of the 
points in Fig. \ref{fig:cycle-comparisons} which lie beyond the circle $I_c = (x^2 + y^2)^{1/2} = 2$ cts/s. The variables $x$ and $y$ refer to the abscissa and ordinate, respectively, of Fig. 6. The points have been binned into 
10 angular bins summed over all intensities.  Figure \ref{fig:residuals-histogram-theta} shows no obvious dependence on angle between 0$^\circ$ and 90$^\circ$ (which is the physically meaningful range) as would otherwise be expected for a positive correlation of the timing of flares in different orbital cycles.  Assuming Poisson errors for each bin, a $\chi^2$ test of the null slope hypothesis between 0$^{\circ}$ and 90$^{\circ}$ confirms this impression: $\chi^2_{\nu} = 1.01$ for a horizontal, straight-line fit. Thus we conclude that there is no statistical correlation between the phasings of flares in one cycle compared to another cycle: 
flare production is apparently a repetitive but stochastic process.

\begin{figure}[htbp]
\begin{center}
\includegraphics[width=6in]{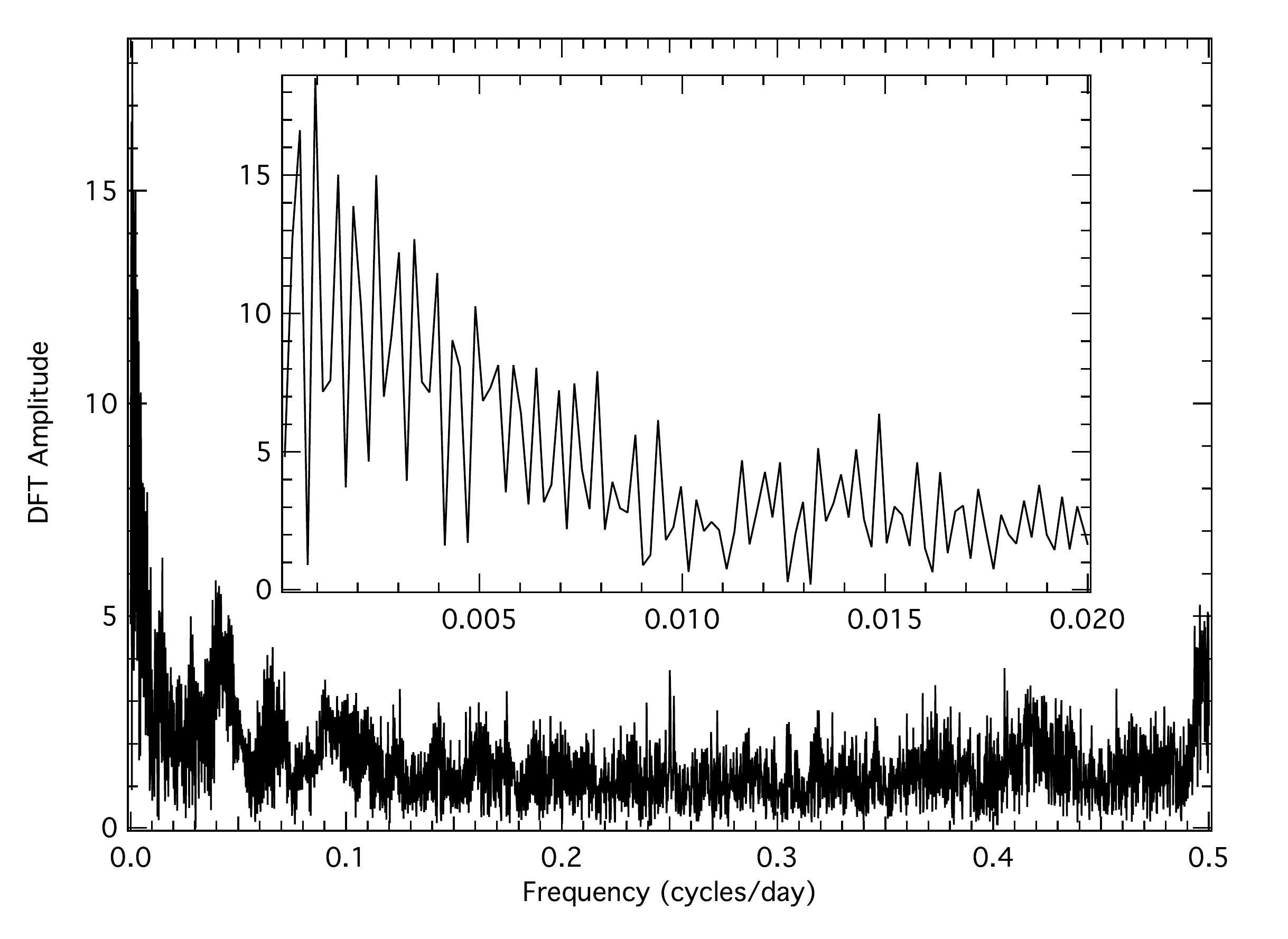}
\caption{Discrete Fourier Transform of residuals (amplitude vs. frequency). Low-frequency amplitude is shown in the inset.}
\label{fig:dft_layout.pdf}
\end{center}
\end{figure}

To further search for 
possible periodicities in the observed flaring events in all three cycles, we Fourier analyzed the whole data string of residual flux using a periodogram appropriate for unevenly sampled data \citep{1982ApJ...263..835S,1986ApJ...302..757H}.  The result is shown in Fig. \ref{fig:dft_layout.pdf}, where the only narrow Fourier amplitude peaks (i.e. strict periodicity) we see are related to \ec's orbital period and its numerous harmonics. Broader flat peaks are seen centered at periods of 67d, 25d, 15d and 11d, suggestive more of the preferred flare time scales rather than any strict periodicity. However, the question remains whether the flares do in fact appear in regularly spaced {\it angular} intervals.  We look at this later in subsection 4.1.2 in the context of one of our suggested models.

\subsection{X-ray Hardness Variations} 

\begin{figure}[htbp]
\begin{center}
\includegraphics[width=6in]{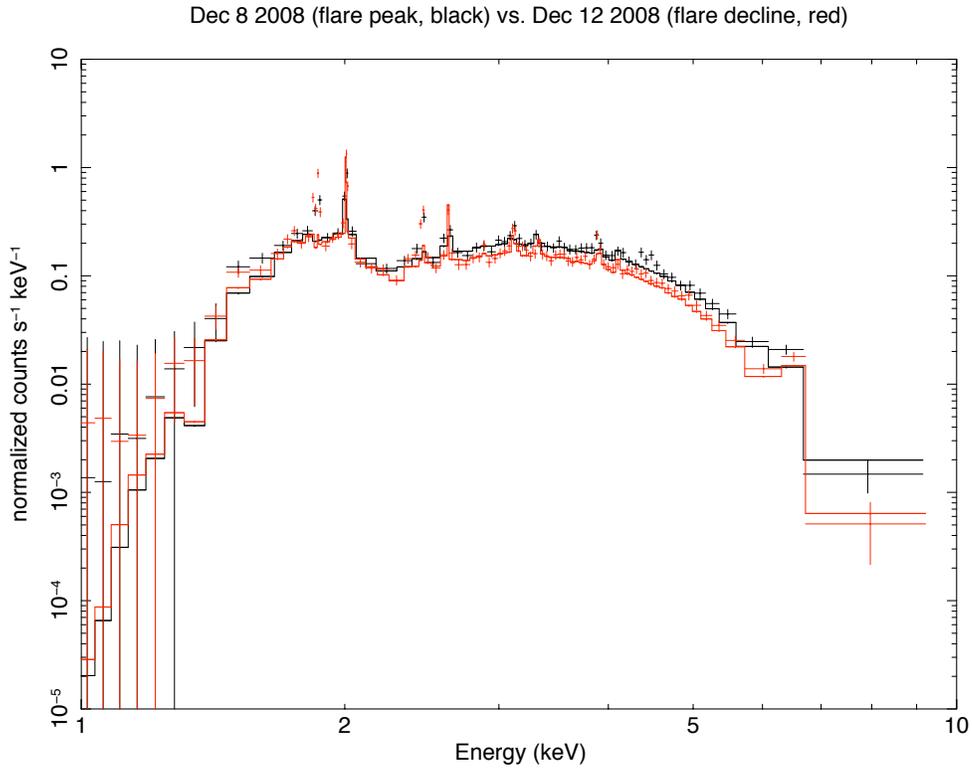}
\caption{Binned \chandra\ MEG $+1$ order X-ray spectrum of the peak of Flare \#43 (black) from Dec 8, 2008 compared to the X-ray spectrum 4 days after the peak, near flare minimum (red). The solid lines are absorbed APEC thermal spectra fit to each spectrum individually.  Note that the single temperature fit does not account for the helium-like Si~XIII triplet near 1.85~keV or the S~XV triplet near 2.45~keV which are prominent in the spectra.}
\label{fig:cmp_10831_10827}
\end{center}
\end{figure}

The hardness ratio analysis presented in Figure 10 of \cite{2005AJ....129.2018C}  shows that some strong flares are associated with significant change in X-ray hardness, in the sense that the flare hardness increases towards the flare peak and then declines.  This impression is confirmed by comparing an \xmm\ X-ray spectrum from 2003 June 13 (near the peak of flare \#23) and another spectrum from an observation on 2003 June 8, prior to the flare rise \citep[see Figure 4 in][]{2007ApJ...663..522H}. These two spectra show that the increase in emission at the flare peak is confined to energies $E \gtrsim 3$ keV for this strong flare.  This suggests an increase in the emission measure of the hottest plasma with little change in low-energy absorption.  Figure \ref{fig:cmp_10831_10827} shows a comparison of \chandra\ spectra obtained on 2008 Dec 8, near the peak of flare \#43, along with a spectrum from an observation on 2008 Dec 12, during the decline of the flare. Once again most of the variation occurs at energies $E > 3$ keV, with little change at lower energies.  

To crudely quantify the spectral change, we fit each spectrum with an absorbed single temperature thermal model using a standard interstellar absorption model \citep{1983ApJ...270..119M} plus a thermal emission plasma model \citep[APEC, ][]{2001ApJ...556L..91S}.  The best fits to the spectra are shown in Figure \ref{fig:cmp_10831_10827}.  The best fit values for $kT$ are $kT=4.9\pm0.5$~keV and $kT=3.6\pm0.5$~keV for the peak and off-peak spectra, respectively.  The best fit columns for the peak and off-peak spectra, respectively, are $N_{H}=5.0\pm0.2$~cm$^{-2}$ and $N_{H}=4.9\pm0.2$~cm$^{-2}$.  Thus the flare peak spectrum can be characterized by a higher temperature than the off-peak spectrum, while there is no significant difference in the absorbing column for either spectrum.

Whatever the mechanism which produces the flares (see \S 4 below), any plausible mechanism leads to a compression and thus additional heating of the shocked gas. This will in turn lead to an increase in temperature and thus X-ray hardness. We describe the compression process more explicitly in subsection 4.1.1.

\subsection{Summary of the Statistical Flare Properties}


The following is a summary of the salient features of the X-ray flares in \ec:

\begin{itemize}
\item   There is a dramatic decrease of flare time-width (FWHM) towards periastron.

\item There is a dramatic increase of flare frequency (or decrease of inter-flare interval) towards periastron. 

\item  The above two effects (time width and time interval) appear to be directly, although not necessarily linearly, correlated.

\item  There is a dramatic increase of maximum flare peak-intensity (H) towards periastron.

\item  Flares appear randomly but on preferred time scales.

\item  During the strong-flare interval just before X-ray minimum, the flare spectra are harder than the non-flare colliding wind emission (and apparently harder than the spectra of flares near the middle of the cycle at apastron).

\item  Within the noise, the maximum value reached for the total strength of the flares (S) is relatively constant with phase.  However, at phases when the flares are narrower, there is a larger spread in observed flare strength.  
\end{itemize}

Any viable model to explain the X-ray flares in \ec\ must take all of these factors into account.  We now look at how to do this in a quantitative way.

\section{Plausible Flare Models}

The temporal behavior and spectral hardness of the flares suggest a disturbance of the interacting region of the two winds where X-rays arise, on random but preferred time scales. The flares must represent a change in X-ray emission level in the relatively small region where the two winds collide nearly head-on, of half opening angle $\theta _X$. This region appears to be significantly smaller in angular size than the overall shock cone opening angle in \ec\ \citep[cf.][]{2008ApJ...680..705H}. The half opening angle of the complete shock cone, $\theta$, can be calculated using the ratio of wind momenta from Table 1: $\eta = (\dot{m_B} V_{\infty,B})/ (\dot{m_A} V_{\infty,A}) = 0.06$ for \ec. From Figure 4 in \cite{1996ApJ...469..729C}, this yields 
the asymptotic opening angle $\theta 
=42^{\circ}$.
For such a wide system, it is unlikely that this angle will be significantly widened by radiative braking from the hotter star B \citep{1997ApJ...475..786G}, until perhaps near periastron when radiative inhibition by star A of the wind from star B could be important \citep{2009MNRAS.tmp..279P}.  In this section, we explore several plausible ways how the flares might occur.  

\subsection{Flares from Structured Winds}

Two distinct kinds of structures 
\citep[e.g.][]{1994Ap&SS.221.....M}
are known to occur in hot-star winds: (a) stochastic multi-scale clumping in all winds and (b) large-scale corotating interaction regions (CIRs, leading to discrete absorption components, DACs, in P Cygni wind line profiles) in many winds. 
We consider below whether the interaction of such structures with the X-ray emitting bow shock could conceivably produce the observed X-ray flares in \ec.

\subsubsection{Flares from Clumped Winds}

Winds from hot stars are believed to be composed of a hierarchy of turbulent clumps, with few large ones and increasingly many more small ones, and with clump size and mass distributed as a power-law \citep[e.g.,][]{2008cihw.conf...17M}. Whether the driver of clumping is some kind of radiative wind instability \citep[eg., ][]{1982ApJ...255..286L,1988ApJ...335..914O} or due to subsurface convection \citep{2009arXiv0903.2049C} remains to be clarified.  In particular, LBVs 
display the greatest amount of stochastic polarimetric variability among all hot luminous stars, due to the large number of clumps in their winds \citep{2005A&A...439.1107D,2007A&A...469.1045D}.  
One conjecture is that the X-ray flares in \ec\ are produced when the rarer, larger clumps enter into the colliding-wind zone, while the background X-ray emission comes from an unresolved superposition of many smaller clumps, which produces similar X-ray emission to that produced by the collision of smooth winds \citep{2007ApJ...660L.141P}. This is analogous to the few discrete emission subpeaks seen on WR or O-star optical emission lines, versus the underlying $\sim$constant wind component from thousands of small wind clumps \citep{1999ApJ...514..909L,1998ApJ...494..799E,2007ApJ...660L.141P, 2008AJ....136..548L}.  In principle 
the clumps could occur in either the wind from star A or star B, since ramming and being rammed lead to essentially the same effect. Assuming the clumps that produce the X-ray flares 
arise in the wind of star A, then the arrival of a big clump from star A into the shock-cone zone will push 
the bow shock towards star B, increasing the density of the shocked portion of star B's wind and thus increasing the overall X-ray emission and hardness for a brief time. Alternatively, the collision of a clump in the wind of either star with the colliding-wind bow shock might produce excess X-ray emission if the clump mixes into the shocked region (a very likely process for adiabatic collisions: Pittard 2007), though the speed of star A's wind (500 km s$^{-1}$) corresponds to temperatures of only $kT \lesssim 0.5$ keV, much lower than the observed flare temperatures ($kT\approx 3-5$ keV).

In any case, 
clumps should expand homologously with the wind. 
Near apastron the colliding wind shock is farther from either star than it is near periastron, and if the clump expands homologously with the wind this could explain qualitatively why the flares are narrower in time (i.e. FWHM) near periastron, i.e. clumps are smaller (and denser) at periastron, 
producing narrow sharp flares near periastron (and, conversely, weaker, longer flares near apastron).  

Below we construct a simple phenomenological model of the interaction of clumps with the X-ray emitting portion of the bow shock. In what follows we assume that the angular size of the X-ray emitting portion of the bow-shock head as seen from either star is $\sim$ constant for all phases.  This is expected if both winds have constant mass-loss rate and constant expansion speed at the point of impact. Near periastron, however, the winds might collide at sub-terminal speeds, depending on the real periastron separation, the wind acceleration (usually parameterized as $V(r)= V_{\infty} (1-R/r)^{\beta}$, 
where $V_{\infty}$ is the terminal velocity, $R$ the stellar radius, $r$ the radial distance and $\beta\approx 1$ for OB star winds), and on whether radiative effects are important.  
We note that recent modeling of the X-ray flux and hardness variation \citep{2008MNRAS.388.1047P} suggests that radiative effects may be important near periastron. 

\begin{figure}[htbp]
\begin{center}
\includegraphics[width=5in]{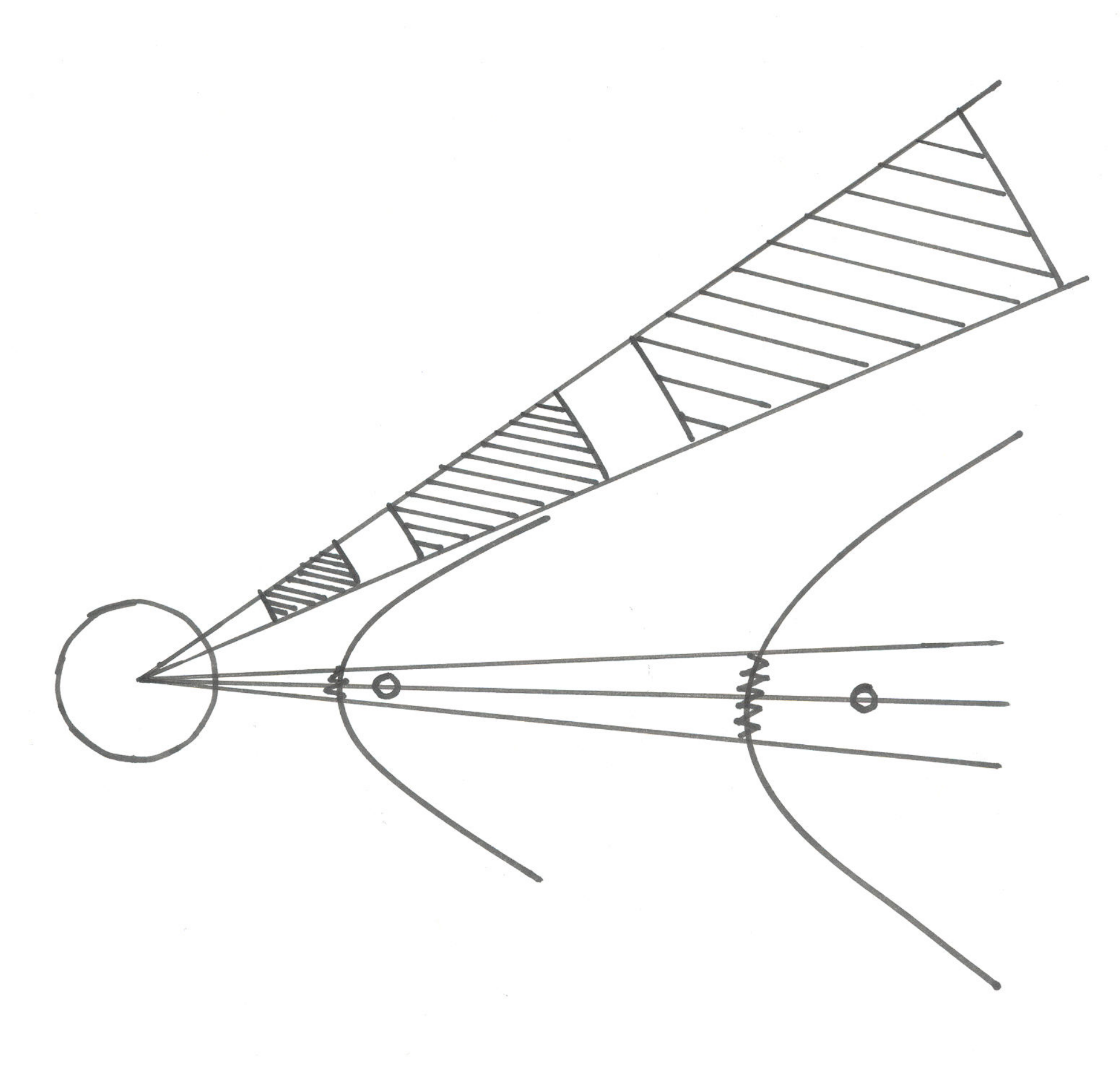}
\caption{Sketch showing a cross section in the orbital plane with 
idealized, outward propagating, homologously expanding clumps of increased radial extent $\Delta r_{cl}$ and of angular width $\Delta \theta \approx 12^{\circ}$ as in Davies et al. (2007)
at three different distances along the same trajectory from star A 
(left), along with the companion (star B) and the associated shock cone, 
at two different separations.  The narrow region of angular width $2 \theta _X$ as seen by star A, where wind-collision X-rays 
arise, is indicated by a zig-zag line at the bow-shock head. Note that $\theta_X << \theta$, since X-rays are produced mostly where the winds
collide nearly head on \citep{2008ApJ...680..705H}. Not to scale.  }
\label{fig:clumpsketch}
\end{center}
\end{figure}

Now we examine the radial clump sizes needed to explain the observed flare width-times. Figure \ref{fig:clumpsketch} shows a 
sketch of this clump model. In simple colliding-wind models, the distance from star A to the stagnation point (where most of the X-ray emission is produced) is $d_A = D/(1 + \eta^{1/2}) = 0.80 D$ while the distance from the companion (star B) to the stagnation point is $d_B = \eta^{1/2} D/(1 + \eta^{1/2})=0.20 D$.  
The orbital separation $D =  a (1-e^2)/(1+e \cos v)$, where $v$ is the orbital true anomaly and $a$ the semi-major axis. At periastron, $D = a (1-e) = 1.54$ AU, and $d_A = 1.23$ AU, $d_B = 0.31$ AU.  At apastron, $D = a (1+e) = 29.3$ AU (19 times larger than at periastron), and $d_A = 23.4$ AU and $d_B = 5.9$ AU.

The shortest observed duration of a flare occurs near periastron, with FWHM $\approx 3$d near $\phi=0.99$ when the separation between the stars is $\approx3$ AU.
%
%
 Assuming this to be similar to the time it takes for a big clump moving radially outward to penetrate the bow shock head and temporarily produce an increase in X-ray flux leads to a clump (radial) size of $\Delta r_{cl} \lesssim \Delta t V_{\infty} \approx 0.9 AU$ for star A and $\lesssim 5.2 AU$ for star B.  
Thus star B cannot be the source of the clumps since the implied  clump size for star B is larger than the distance from star B to the stagnation point at $\phi=0.99$ ($\approx 0.6$ AU), unless the clump speed is much less than the terminal velocity of the wind of star B (which seems unlikely). 
For star A, there 
is sufficient (although not ample) room at periastron (and even more room in an absolute sense at other times); with a lower wind speed close to the LBV with a radius of $\approx$ 0.3 AU, 
there is even more room. However one expects a clump moving at $\sim500$ km s$^{-1}$ to produce shocked gas with a temperature of $\lesssim 10^{6}$ K, not the $\approx 3\times10^{7}$ K emission seen by \rxte\ during the flares.  Nevertheless, we assume that star A is the source of the clumps, and that 
the excess density of the clump becomes mixed into the hotter shocked gas in the unstable region near the wind-wind collision \citep{2007ApJ...660L.141P}, or that as the clumps impinge on the bow shock they move the bow shock into denser regions of star B's hotter wind.

What kind of 
X-ray luminosity enhancement can be expected when a clump traveling at the terminal wind speed from star A pushes into the WWC zone? Assume that a large clump from star A's wind 
enters and dominates the WWC zone, but with density $\rho'_{A}>\rho_{A}$, where $\rho_{A}$ is the background density of star A's wind. 
Then, assuming mass conservation, the corresponding wind-momentum ratio within the WWC zone where X-rays arise will be modified temporarily to $\eta' = \eta/\mu$, where $\mu = \rho'_{A}/\rho_{A}$. This will then 
push the stagnation point closer to star B, since $d_B = D \eta'^{1/2}/(1 +  \eta'^{1/2})=D(\eta/\mu)^{1/2}/(1+(\eta/\mu)^{1/2})$.  Then, assuming $L_X \propto 1/d_B$, one has $L_X^{'}/L_X = (\mu^{1/2} + \eta^{1/2})/(1 + \eta^{1/2}) = 1.8$ for the largest flares (as observed), assuming $\mu=4$ as expected for strong shocks. 

Neglecting tangential motions of the bow shock and the clumps for simplicity, the flare time-width can be approximated by: $\Delta t (FWHM) \approx \Delta r_{cl}(d_A)/V(d_A)$, where $\Delta r_{cl}(d_{A})$ is the radial extent of the clump at the distance of the shock-cone head $d_A$ from star A.  
Then, homologous clump growth in the wind yields $\Delta r_{cl}(d_A) = \Delta r_{cl}(R_A)d_A/R_A$, where $\Delta r_{cl}(R_A)$ is the clump size near the sonic point of star A.
Thus, 
\begin{equation}
\Delta t (FWHM) = k \times D/(1 - R'_{A}/D)^{\beta}, 
\label{eq:clumpfwhm}
\end{equation}
with constant $k = \Delta r_{cl}(R_A)/(V_{\infty,A} R'_{A})$ 
and $R'_{A} = (1 + \sqrt{\eta}) R_{A}$. If $\beta=0$, then  $V = V_{\infty,A}$ and the flare durations (i.e. time-widths) are $(1+e)/(1-e) = 19 \times$ shorter at periastron than at apastron.  In Fig. \ref{fig:fwhm} we show a least squares fit of equation \ref{eq:clumpfwhm} to the observed flare FWHMs (assuming $\beta = 1$).  Given the more uniform dispersion of the $\Delta t (FWHM)$ values in log than in linear units, we optimized the fit in log space.  

Using the parameters from Table 1 along with $\eta=0.06$ and $\beta=1.0$ (Hillier et al. 2001), the best-fit clump model provides a good description of the observed FWHMs with $k=1.7 \pm 0.1$ d/AU. 
%
Using this value of $k$, $\Delta r_{cl}(R_A) \approx 0.17 AU \approx 0.61R_{A}$.  This implies that the largest clumps occupy more than half the stellar radius in the radial direction near the sonic point, if indeed they actually start there, suggesting that the clumps are radially extended.
At apastron the distance to the bowshock head is $d_{A}=D(\mbox{apastron})/(1+\sqrt{\eta}) =23.5$AU, and the clump size in the radial direction is still the same fraction of the distance according to homologous expansion, i.e. $\Delta r_{cl}\approx 0.61 d_{A} \approx  14$AU, nearly equal to the adopted size of the semi-major axis. 
We also found from eq. 2 a spread in $\Delta t (FWHM)$ from 3.4d at periastron to 50d at apastron. Since the time of periastron passage $T_o$ was assumed to coincide with the beginning of the X-ray eclipse, we refitted eq. \ref{eq:clumpfwhm} to the data  allowing $T_o$ to vary.  We found that a slightly earlier value of $T_o$ by $7 \pm 4$d fits but this fit is only marginally better. Since there is little evidence at other wavelengths that periastron occurs before the X-ray minimum, we therefore ignore this and retain the assumption that periastron passage occurs at the start of the X-ray minimum.   

On the other hand, assuming that (1) clumps form stochastically but on a quasi-continuous and constant basis per unit time and in all directions around \ec, (2) they expand truly homologously in all three dimensions (radial and two transverse), and (3) they collide in a constant colliding-wind zone angle, there is no reason to expect the separation time between successive flares, $\Delta t(sep)$, to vary around the orbit.  This is contrary to what is seen in 
Fig. \ref{fig:tm_flare_fwhm_ave_sep}, however, where the variation in average flare time-separation is clearly correlated with the variation in FWHM.  

The correlation between the observed flare FWHMs and 
true separation 
may be
a consequence of reduced detectability of clumps towards apastron.  Quantitatively, this can be explained as follows.  
Flares are only detected when the residual flare height $H > H_c \approx 1$~ct~s$^{-1}$, the threshold value 
derived above.  
We further assume that the total strength of a flare $S$ follows a power law, $N(S) dS = N_o S^{-\alpha} dS$, where $\alpha$ and $N_{o}$ are constants. This is inspired by the observed power law in Fig. 7 and similar behavior for sub-peak spectral intensities from clumps in WR winds \citep{1994RvMA....7...51M}.    Then at a given orbital phase $\phi$, where $W \equiv FWHM$ takes on a preferred value, 
we can only detect flares with $S > W H_c$. 
This leads to the total number of flares detected per unit time 
$$N_{tot}(W) = \int_{WH_c}^{S_{max}} N(S) dS = \frac{N_{o}}{(1-\alpha)}[S_{max}^{1-\alpha} - (WH_c)^{1-\alpha}],$$ 
%
where $S_{max}$  is the maximum value of $S$ from the largest clumps. 
(Note that once absorbed into the adiabatic WWC zone, hydrodynamic models predict that the clumps will be quickly destroyed, i.e. merged into the ambient material: Pittard 2007.) Then the mean separation between successive detected flares is 
%
%
%
%
$$\Delta t (sep) = 1/N_{tot}(W) = k_{2} W^{\alpha -1}/(k_{1} - W^{\alpha -1}),$$
with constants $k_1 = (S_{max}/H_c)^{\alpha - 1}$ and $k_2 = (\alpha -1) S_{max}^{\alpha -1}/N_o$.  Reversing this relation, we can also write 
\begin{equation}
W = [k_1 \Delta t(sep)/(k_2 + \Delta t(sep))]^{1/(\alpha -1)}.
\end{equation}

A log-log fit of this last relation (with $W$ and $\Delta t$ in days; more mathematically stable than the above equivalent) is shown in Fig. 4, with $\chi^2$ 
- minimized parameters $\alpha = 1.7 \pm 0.2, k_1 = 23 \pm 24, k_2 = 69 \pm 52.$  While $k_1$ and $k_2$ are poorly constrained (although necessary to characterize the relation), the power-law index $\alpha$ is very close to that found for WR clumps: $\alpha = 2.0 \pm 0.2$ (Moffat 1994). 
From Figure \ref{fig:strength} we find that $S_{max}\approx 10^{7}$ counts, and, with $k_1^{1/(\alpha -1)} = 88d$, this yields $H_c \approx 10^7 /k_1^{1/(\alpha -1)} =$ 1.3 cts/sec, very close to that above.

Overall, the fact that the clump model fits so well in both $\Delta t(FWHM)$ and in $\Delta t (sep)$ implies that the actual rate of clump production must indeed be uniform with time and the clumps do survive at least from 1.2 to 23 AU, the separation of the bow-shock head from periastron to apastron. We can now calculate the total number of observable flares at periastron and extrapolate to the whole wind to estimate the total number of clumps.  Taking $W = 3d$ near periastron, we find $N_{tot}(W) = (k_1 - W^{0.7})/(k_2 W^{0.7}) \approx $ 0.14 flares/d, or 0.49 flares/3.5d, i.e. during one wind-expansion time $R_A/V_{\infty}$. If one flare corresponds to one clump of angular diameter $\Delta \theta ^{\circ}$ and clumps are distributed uniformly in a spherically symmetric wind around the star, then we find $0.49 \times 4.1 \times 10^4 /(\Delta \theta ^{\circ})^2 =$ 2300, 205, 23 clumps per wind expansion-time for $\Delta \theta ^{\circ} =$ 3, 10 and 30, respectively.  If $\Delta \theta = 3^{\circ}$, as in WR winds (\citealp{2002A&A...393..991D}; but see \citealp{2003A&A...406L...1D})
then the first value is most likely, while Davies et al. (2007) find $\Delta \theta ^{\circ} = 12^{\circ}$ for LBV winds.  These are compatible with the total number ($\sim 10^3$) of clumps found in the inner parts of LBV winds by \cite{2007A&A...469.1045D} 
based on polarimetry.

\subsubsection{Flares from Large-scale Co-Rotating Interaction Regions}

Winds from hot stars are also widely observed to have rather stable structures \citep[spiral density waves or ``co-rotating interaction regions'', CIRs,][]{1986A&A...165..157M} which are apparently tied in some way to the stellar surface and which are swept around with the star as it rotates.  Such structures produce ``Discrete Absorption Components'' (DACs) observable in the absorption troughs of unsaturated P Cygni wind lines. One possibility for the production of the X-ray flares is that the orbital motion of the companion carries it through such a stable density enhancement in the wind, which is itself swept around by the rotation of the star (see sketch in Fig. 12). 
The mechanical effect of this interaction would be the same as for clumps: this interaction of the bow shock with a stable density enhancement could push the bow-shock towards the companion, to regions of higher density in the companion's wind, thus increasing the emission measure of the hot shocked gas. The emission measure enhancement would only last as long as it takes for the relative motion of the companion to carry it through the CIR. 

If the detectability of CIR passages suffers from the same problem as that 
for the clumps (i.e. slower sweeping at apastron, making for 
lower-intensity, broader flares, many of which are more likely to remain 
below the noise threshold), then the rate of sweep must be determined when 
they are best seen, i.e. at periastron, where $\Delta t (sep) < $10 d (cf. Table 2; if detectability is still a problem even at periastron, then this limit 
could be decreased even more).  Then for $n$ CIRs distributed more or less 
evenly around star A's equatorial plane (which is assumed to be aligned with the 
orbital plane for simplicity), the rate of sweep will be $d \theta_{CIR}/dt > (2 \pi/n)/$10 rad/d.  For 
the most plausible number of CIRs, $n =$ 2 
\citep{1999A&A...344..231K}
this yields  
$d\theta_{CIR}/dt >$ 114 rad/yr.

Now, the net sweep rate of CIRs as seen by the companion (and the shock 
zone) is $d \theta_{CIR}/dt = dv/dt - \Omega$, where
$dv/dt$ is the sweep rate due to orbital motion (varying from 0.137 rad/yr at 
apastron to 49.4 rad/yr at periastron) 
given by
$$\frac{d\upsilon}{dt}=\frac{2 \pi}{P} \frac{(1 + e \cos\upsilon )^{2}}{(1 - e^2)^{3/2}}$$
and $\Omega$ is the stellar rotation 
rate seen by a distant inertial observer.  
%

At periastron, we have $\Omega <$ 49.4 - 
113.9 = -64.5 rad/yr.  This can be compared with the critical rotation rate 
of star A: $\Omega_c = \sqrt{GM_A/R_A^3} =$ 60.0 rad/yr.  This means that, in 
absolute terms, the star would have to be rotating faster than critical.  This problem can 
be reduced by choosing $n > 2$. 
But
the larger $n$ is, the 
more $d \theta_{CIR}/dt$ will be dominated by orbital rotation, which is highly 
non-uniform around the orbit, removing the need for variable detectability 
(see below).  
There are other problems with this model, however.  For example
for small 
$n$, $\Omega$ is in a direction opposed to that of the orbit. 
In addition this model only works if the orbital and rotational plane of star A are co-planer, which may be unlikely for such a wide binary.

The rotation of star A will necessarily impose curvature on any CIR originating near the surface of the star (e.g. Cranmer \& Owocki 1996). However, to explain the observed strongly varying time-separation around the orbit 
requires $dv/dt$ to dominate over $|\Omega|$ at most phases.  Because of this, and for the sake of simplicity, we ignore curvature of the CIRs in what follows.  Then the variation of the flare FWHM and the average flare separation as a function of time can be expressed as follows:
\begin{equation}
\Delta t(FWHM) = (2 \theta_{X})/(d\upsilon/dt-\Omega) = \frac{P\theta_{X}(1 - e^2)^{3/2}}{\pi(1 + e \cos\upsilon)^{2}-(1 - e^2)^{3/2} P \Omega /2}
\label{eq:fwhmcir}
\end{equation}
%
%
%
and 
\begin{equation}
\Delta t(sep)= \frac{2\pi}{n}\frac{1}{(dv/dt-\Omega)} =  \frac{\pi}{n \theta_{X}} \Delta t(FWHM).
\label{eq:sepcir}
\end{equation}
%
The CIR model 
directly predicts a linear correlation between the FWHMs and flare separations.  
Fitting eq. \ref{eq:fwhmcir} to the observed FWHM vs phase yields $\theta_{X}=0.08$ radians or $\theta_{X}=4^{\circ}$ with a value of $\Omega = 0.014$ days$^{-1}$ or $P_{rot}=450$ days.  We then constrained $\theta_{X}$ between 0.06 and 0.1 radians, and fit eq. \ref{eq:sepcir} to the observed FWHM vs. average flare separation. This yielded $n = 75$ co-rotating interaction regions around the star.  This value of $n$ is far higher than is normally encountered in OB stars, while $\theta_{X}$ is relatively small. In a sense, such a large value of $n$ more closely matches the notion of a large number of clumps. 
Furthermore, allowing for 
curvature, which is stronger for larger $\Omega$, also introduces an asymmetry in $\Delta t (FWHM)$ or $\Delta t (sep)$ vs. phase, 
depending on whether one is receding from or approaching periastron, which 
is not seen.
Overall, then, it appears difficult 
to explain $\eta$ Car's X-ray flares via CIR action.

\subsection{Flares from Intrinsic Instabilities in the Colliding-wind Region}

Numerical smoothed particle hydrodynamics (SPH) models of the X-ray emission produce by the wind-wind collision in \ec\ by \cite{2008MNRAS.388L..39O} shows ``flare-like'' instabilities especially  prominent just before periastron.  The behavior of the SPH models is qualitatively similar to the observed behavior of the \rxte\ X-ray lightcurve. 
These variations in the model lightcurves were due to transient absorption of the X-ray emission (assumed to originate from a point-like region near the stagnation point) by blobs produced near the unstable wind-wind interaction surface passing through the observer's line of sight.  As discussed above the X-ray spectra from a number of flares shows that the flares instead result from excess high energy emission rather than increases in absorption.  But the wind-wind interaction in \ec\ may be radiatively unstable near periastron \citep{2002AA...383..636P,2009MNRAS.tmp..279P}, and it's worth considering whether such instabilities could produce the flare-like behavior we observe.

Any gas with density $\rho$ has a preferred timescale for oscillations, which goes as the free-fall time: $\delta t \sim t_{ff} = 0.54/(G\rho)^{1/2}$.  Assuming that the density in the bow-shock head where the excess X-rays arise is proportional to the arriving wind density at the point of impact, and with $\dot{M} = 4\pi r^2 \rho(r) V(r)$, one finds $\delta t \propto r V(r)^{1/2}$. Thus, for $V(r) \sim V_{\infty}$ and constant for most of the orbit, $\delta t \propto r$, as observed for $\Delta t (FWHM)$.   All this may explain the time scale variations (widths of flares). 
But density variations produced by stochastic instabilities should show both overdensities and underdensities, which would produced bright peaks and fainter troughs, respectively.   While we see clear evidence of brightness enhancements, there is little evidence for such troughs in the \rxte\ X-ray lightcurve, especially away from the X-ray minimum in the interval $0.2<\phi<0.8$, where there is less crowding and troughs should be more obvious if they existed.  Also, the observed excess peaks are much narrower than any possible deficit, contrary to what is expected if due to oscillations.
Because of these difficulties, the whole idea of oscillations does not seem too convincing, although more detailed considerations may be 
useful.  Such considerations require detailed numerical models which are 
beyond the scope of this paper.   


\section{Conclusions}

Basically, all three models have varying degrees of problems.  In addition, there is the shared problem that the orbital parameters of \ec\ are not very well known. But despite the inherent subjective problem of actually defining the flares in \ec\, we believe that the clump model is the simplest (and therefore most likely) explanation yet found for the flares.  LBV stellar pulsations on a time scale of $\sim$85 d seem less likely because of the lack of a true oscillation signature.

Clumping on relatively large scales has been invoked to explain the
X-ray lines in O stars \citep{2003A&A...403..217F}. \cite{2007A&A...476..335W}
use similar model parmeters to explain flaring in X-ray
binaries. Thus, wind collisions have some similarity to massive X-ray
binaries, in which relatively hard X-ray flares are seen as a likely result
of clumpy winds being accreted by the compact component.  If so, then
one should should see X-ray flares in other wind-wind colliding systems, 
such as the prototype WR + O system WR140 = HD 193793, WC7pd + O5, P = 7.94 yr, e = 0.88.  The X-ray lightcurve of WR140 is also being monitored by \rxte. If it shows flares (in fact there is some evidence for absorption dips possibly from clumps), then they are far less obvious than those in \ec.  One difference between these two systems, though, is that the two colliding winds in WR140 are both very fast and may not contain the 
prominent clumps that LBV winds appear to have \citep{2007A&A...469.1045D}, 
leading to the giant X-ray flares that make \ec\ so spectacular.




\acknowledgments
AFJM is grateful to NSERC (Canada) and FQRNT (Qu\'ebec) for financial support. MFC gratefully acknowledges NASA for the support of the ongoing \rxte\ observing campaign, and from \chandra\ grant GO7-8076A. This research has made use of data obtained from the High Energy Astrophysics Science Archive Research Center (HEASARC), provided by NASA's Goddard Space Flight Center. This research has made use of NASA's Astrophysics Data System. We also acknowledge the efforts of the \rxte\ Science Operations Facility for help in scheduling these observations. We thank the anonymous referee for helpful suggestions.



\appendix

\section{Flare Properties}

\begin{deluxetable}{rccrrr}
\tablewidth{0pt}
\tablecaption{Flare Parameters}
\tablehead{
\colhead{ID} & 
\colhead{Flare Centroid} & 
\colhead{Orbital Phase} & 
\colhead{FWHM} & 
\colhead{Peak Height ($H$)} &
\colhead{Ave. Separation} \\
\colhead{} & 
\colhead{Years} & 
\colhead{} & 
\colhead{Days} & 
\colhead{Counts s$^{-1}$} &
\colhead{Days} }
\startdata
       -11 &    1996.57 &      0.749 &       31.8 &       1.43 &        ---    \\
        -9\tablenotemark{a} &    1997.06 &      0.837 &       33.8 &       1.67 &      127.8 \\

        -8 &    1997.27 &      0.875 &       35.7 &       2.39 &       71.2 \\

        -7 &    1997.45 &      0.908 &       29.8 &       2.15 &       60.3 \\

        -5\tablenotemark{a} &    1997.60 &      0.935 &       23.8 &       3.82 &       42.0 \\

        -4 &    1997.68 &      0.949 &       15.9 &       3.35 &       23.7 \\

        -3 &    1997.73 &      0.958 &       19.8 &       7.53 &       23.7 \\

        -2 &    1997.81 &      0.973 &       14.7 &       5.02 &       25.6 \\

        -1 &    1997.87 &      0.983 &       20.2 &      11.47 &       21.9 \\

         0 &    1997.93 &      0.994 &        6.7 &       3.82 &       X-ray Minimum\tablenotemark{b}      \\

         1 &    1998.20 &      0.043 &       19.7 &       2.39 &       X-ray Minimum      \\

         2 &    1998.47 &      0.092 &       28.3 &       1.55 &      115.1 \\

         3 &    1998.83 &      0.157 &       34.9 &       1.43 &      105.9 \\

         4 &    1999.05 &      0.196 &       63.5 &       1.67 &      142.4 \\

         5 &    1999.61 &      0.297 &       47.6 &       1.19 &      162.5 \\

         6 &    1999.94 &      0.357 &       31.8 &       0.96 &      160.7 \\

        7a\tablenotemark{a} &    2000.49 &      0.456 &       43.7 &       0.96 &      133.3 \\

        7b &    2000.67 &      0.489 &       51.6 &       1.43 &      124.2 \\

         8 &    2001.17 &      0.579 &       35.7 &       1.43 &      169.8 \\

         9 &    2001.60 &      0.657 &       51.6 &       1.55 &      129.7 \\

        10 &    2001.88 &      0.707 &       59.6 &       1.19 &      129.7 \\

        11 &    2002.31 &      0.785 &       33.8 &       2.99 &      127.8 \\

        12 &    2002.58 &      0.833 &       19.8 &       2.87 &       69.4 \\

        13 &    2002.69 &      0.853 &       23.8 &       1.19 &       53.0 \\

        14 &    2002.87 &      0.886 &       21.8 &       2.03 &       51.1 \\

        15 &    2002.97 &      0.904 &       19.8 &       1.67 &       34.7 \\

        16 &    2003.06 &      0.920 &       19.8 &       2.87 &       38.4 \\

        17 &    2003.18 &      0.942 &       21.8 &       1.67 &       38.4 \\

        18 &    2003.27 &      0.958 &       13.9 &       5.25 &       27.4 \\

        19 &    2003.33 &      0.969 &        6.7 &      10.00 &       20.1 \\

        20 &    2003.38 &      0.978 &        6.0 &       3.82 &       12.8 \\

        21 &    2003.40 &      0.981 &        5.2 &      10.00 &        9.1 \\

        22 &    2003.43 &      0.987 &        2.8 &       4.78 &       11.0 \\

        23 &    2003.46 &      0.992 &        4.8 &      14.80 &        X-ray Minimum    \\

        24 &    2003.80 &      0.054 &       39.7 &       1.43 &        X-ray Minimum     \\

        25 &    2004.23 &      0.131 &       31.8 &       0.74 &      102.3 \\

        26 &    2004.36 &      0.155 &       35.7 &       0.93 &       60.3 \\

        27 &    2004.56 &      0.191 &       43.7 &       0.93 &      204.5 \\

        28 &    2005.48 &      0.357 &       71.5 &       1.07 &      244.7 \\

        29 &    2005.90 &      0.433 &       91.3 &       0.60 &      162.5 \\

       30a\tablenotemark{a} &    2006.37 &      0.517 &       63.5 &       1.91 &      122.4 \\

       30b &    2006.57 &      0.554 &       83.4 &       1.67 &      102.3 \\

        31 &    2006.93 &      0.619 &       59.6 &       1.43 &      129.7 \\

        32 &    2007.28 &      0.682 &       40.8 &       1.07 &       98.6 \\

        33 &    2007.47 &      0.716 &       49.0 &       0.94 &       85.8 \\

        34 &    2007.75 &      0.767 &       49.0 &       1.88 &      100.4 \\

        35 &    2008.02 &      0.815 &       49.0 &       1.61 &       73.1 \\

        36 &    2008.15 &      0.839 &       40.8 &       1.07 &       54.8 \\

        37 &    2008.32 &      0.869 &       31.0 &       1.88 &       53.0 \\

        38 &    2008.44 &      0.891 &       20.4 &       1.15 &       36.5 \\

        39 &    2008.52 &      0.905 &       22.1 &       0.94 &       58.4 \\

        40 &    2008.76 &      0.949 &       11.7 &       3.33 &       51.1 \\

        41 &    2008.80 &      0.956 &        7.3 &       2.82 &       14.6 \\

        42 &    2008.84 &      0.963 &        9.1 &       7.42 &       23.7 \\

        43 &    2008.93 &      0.979 &       13.8 &      15.75 &       23.7 \\

        44 &    2008.97 &      0.987 &        6.0 &       9.00 &       12.8 \\

        45 &    2009.00 &      0.992 &        2.2 &       7.75 &       X-ray Minimum      \\
\enddata   
\tablenotetext{a}{Some gaps or additions occur in the flare numbering due to omission after initial assignment too close to the noise level or two originally selected flares (nos. 7 \& 30) being subsequently split into two for greater consistency, while no. 8 was left as one.}
\tablenotetext{b}{Flares are not observed during the X-ray minima, so we do not calculate the average separation for the last flare before the minimum or the first flare after minimum.}
\label{tab:flareparams}
\end{deluxetable}

\bibliography{flares}


















\clearpage


\clearpage 

\clearpage

\clearpage

\clearpage

\clearpage

\clearpage

\clearpage

\clearpage

\clearpage

\clearpage

\begin{figure}[htbp]
\begin{center}
\includegraphics[width=5in]{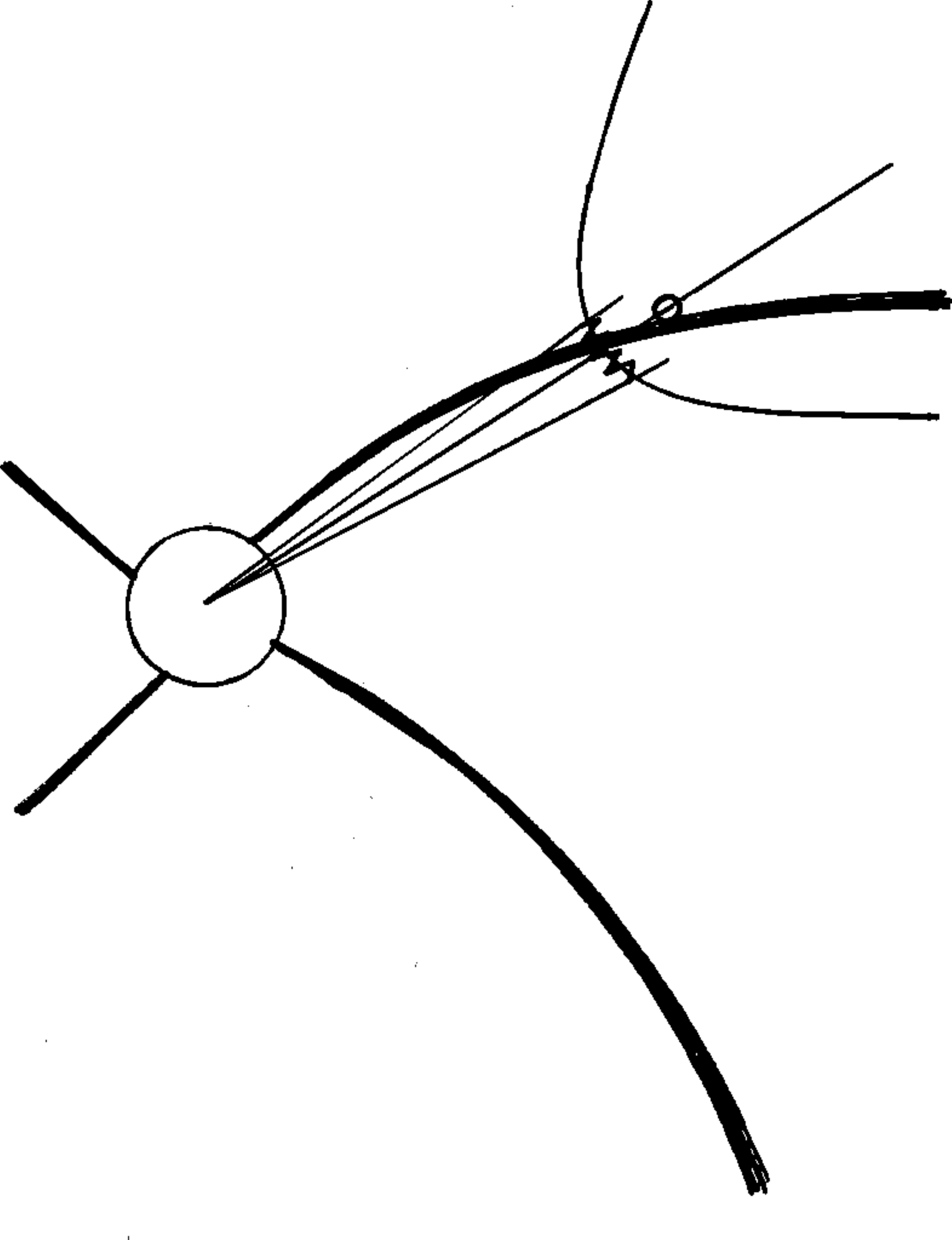}
\caption{Cartoon showing a cross section in the orbital plane with four 
idealized CIRs rotating counterclockwise with \ec\ A (left).  One of the 
CIRs is shown entering (or being entered by - depending on the relative 
angular speed of each) the narrow region at the orbiting bow shock head (angular size $2 \theta_{X}$)
where X-rays arise. Not to scale.}
\label{fig:cirsketch}
\end{center}
\end{figure}

\end{document}